\documentstyle[12pt,epsfig]{article}
 1
 1
 1

\newcommand{\beq}{\begin{equation}}
\newcommand{\eeq}{\end{equation}}
\newcommand{\beqn}{\begin{eqnarray}}
\newcommand{\eeqn}{\end{eqnarray}}
\newcommand{\ra}{\rightarrow}

\newcommand{\gag}{$\gamma \gamma$ }
\newcommand{\bruit}{S/\sqrt{B}}
 
\newcommand{\la}{\lambda}

\newcommand{\lambdae}{\lambda_{e}}

\newcommand{\np}{{\em Nucl.\ Phys.\ }}
\newcommand{\pl}{{\em Phys.\ Lett.\ }}
\newcommand{\pr}{{\em Phys.\ Rev.\ }}
\newcommand{\prl}{{\em Phys.\ Rev.\,Lett.\ }}

\newcommand{\Mhe}{M_{H}}

\newcommand{\she}{\hat{s}}

\newcommand{\cl}{{{\cal L}}}

\newcommand{\bbbar}{b\overline{b}}
\newcommand{\ccbar}{c\overline{c}}
\newcommand{\qqbar}{q\overline{q}}

\def\cpviol{${\cal{C}} {\cal{P}}$}
\def\slashc{c\kern -.400em {/}}
\def\slashL{L\kern -.450em {/}}
\def\slashcl{\cl\kern -.600em {/}}
\def\gam{\gamma \gamma}

\def\gag{$\gamma \gamma$}

\def\sm{${\cal{S}} {\cal{M}}$}

\def\epem{\ifmmode{{e^{+}e^{-}}}\else{${e^{+}e^{-}}$}\fi}
\newcommand{\epm}{$e^{+} e^{-}$}

\begin{document}
\bibliographystyle{unsrt}
\def\baselinestretch{1.2}

\relax

\begin{titlepage}
\rightline{ENSLAPP-A-485/94}
\rightline{UdeM-GPP-TH-94-9}
\rightline{hep-ph/9409263}
\rightline{September 1994}
\vspace*{\fill}
\begin{center}
{\large 
{\bf
   Extracting  the Intermediate-Mass Higgs Resonance 
             at Photon-Photon Colliders}}
\vspace*{0.5cm}

\begin{tabular}[t]{c}
M. Baillargeon$^{1,2}$, G. B\'elanger$^{1,2}$ and F. 
Boudjema$^{1}$\\
{\it 1.  Laboratoire de Physique Th\'eorique 
EN{\large S}{\Large L}{\large A}PP} $^{\S}$\\
{\it B.P.110, 74941 Annecy-Le-Vieux Cedex, France}\\
{\it 2. Laboratoire de Physique 
Nucl\'eaire, Universit\'e de Montr\'eal},\\
{\it  C.P. 6128, Succ. Centre-Ville, Montr\'eal, 
Qu\'ebec, H3C 3J7, Canada.} \\
\end{tabular}
\end{center}
\vspace*{\fill}

\centerline{ {\bf Abstract} }
\baselineskip=14pt
\noindent
 {\small
   We address the problem of how to extract the signal of a Higgs within the 
intermediate mass range at a photon-photon collider that has a wide 
energy spectrum. 
All backgrounds from two-jets production are included: direct, 
so-called resolved and twice-resolved as well as single $Z$ and $W$ 
production. Uncertainties in the evaluation of the QCD-initiated processes 
such as the choice of structure function and the issue of 
radiative corrections are discussed. 
We consider various combinations 
of the polarizations and invariant mass resolutions as well as
jet-tagging strategies with different efficiencies. 
The analysis is based on 
an automatized technique that, given a specific detector and machine 
configuration, 
returns the 
optimal set of cuts corresponding to the best significance one may hope 
to achieve for each particular Higgs mass. 
We find that at a photon machine obtained from a 500~GeV 
\epm\ linear collider with $\int\!{\cal L}=10$~fb$^{-1}$ it will be 
possible to extract a Higgs signal in the range 110--140~GeV,
while with the same luminosity, a 350~GeV option not only 
extends the discovery limit down to 90~GeV but gives much better significance levels. 
}

\vspace*{\fill}

 
\vspace*{0.1cm}

\noindent 
{\footnotesize $\;^{\S}$ URA 14-36 du CNRS, associ\'ee \`a l'E.N.S. 
de Lyon et au LAPP 
d'Annecy-le-Vieux.}

\end{titlepage}
\baselineskip=18pt

\setcounter{section}{1}
\setcounter{subsection}{0}
\setcounter{equation}{0}
\def\thesubsection {\thesection.\arabic{subsection}}
\def\theequation{\thesection.\arabic{equation}}

\setcounter{equation}{0}
\def\thequation{\thesection.\arabic{equation}}

\setcounter{section}{0} 
\setcounter{subsection}{0}
\section{Introduction}

Given the  evidence for top production 
claimed by the CDF collaboration\cite{CDFtop},
with a mass that is well in accord with 
the indirect limits by LEP/SLC\cite{toplep}, all the entries that make up the 
particle content of the standard model, \sm, would be filled were it not for 
the notable exception of the Higgs. 
Yet the discovery of the very elusive Higgs particle is crucial to 
our present understanding of the important mechanism of symmetry breaking and
the concomitant problem of the origin of mass and \cpviol violation. 
No wonder then, that the 
Higgs searches (or aspects intimately related to them) are one of the most 
prominent motivations in almost 
all proposals for new high-energy physics machines. 

At the moment all we know 
is that, if the Higgs exists, it is heavier than $64.5$~GeV\cite{Glasgow94}.
 Even with the 
fantastic precision reached by LEP and SLC, combined with low-energy 
data, no firm conclusion about the mass of this particle  can be extracted 
from its quantum effects\cite{toplep}.
We will have to wait for LEPII which, if operated with the design energy, 
$190$~GeV, and luminosity, $\int \!\cl=500$~pb$^{-1}$, would 
cover the mass range up to $90$~GeV\@. The next hope will be the LHC.
Unfortunately, 
the hadron machine will only be able to efficiently cover a Higgs with a 
mass in 
excess of $140$~GeV\@. The ``mass-gap" 90--140~GeV, that has come to be 
known as the intermediate mass Higgs, IMH, will be very arduous to cover at 
this machine since the Higgs will decay predominantly into $b\bar b$. 
Even  
with excellent methods of $b$  recognition and rejection of other
flavours, the $pp$ environment is not 
conducive to such a search. {\it Faute de mieux}, 
there is of course the possibility of trying 
to hunt the IMH through its two photon decay.
However, to discover a Higgs via this decay mode is a
very difficult task that requires a dedicated 
expensive detector tailored for it.
The intermediate mass Higgs is, on the other hand, an important and 
quite special eventuality. Naturalness arguments, as exemplified 
by supersymmetry, do require a light Higgs, and therefore it is of utmost 
urgency to cover the ``mass-gap". In this respect, an \epm\ collider
such  as the 
much discussed NLC (Next Linear Collider), which in a first phase is planned to
be operated at a centre-of-mass energy between 350~GeV and 500~GeV, will not miss this
Higgs\cite{DesyNLC}.

One very attractive option for the search of the IMH 
is a photon-photon collider\cite{PhotonCol}
obtained from backscattering laser light on the beam of a high energy linear
electron collider. This is  all the more interesting since the missing 
fundamental particle of the \sm\ is the only one which is spin-less 
and could thus  couple to two photons. Therefore, 
an intense and high energy \gag\  collider has
the quite unique capability of 
 producing a scalar particle as a resonance.
This  resonant Higgs structure is out of reach in the usual \epm\ mode since  
chirality highly suppresses this $s$-channel production.
The drawback of the photon collider 
is that the Higgs, being neutral, couples to two 
photons only at the loop-level. The rate 
of production is therefore not so large, and the resonant structure would
not be as prominent as, for example, 
the beautiful spin-1 $Z$ peak  one observes at 
LEP. 
The one-loop $\gamma\gamma$ initiated Higgs 
production mechanism  is nevertheless an interesting feature, since
a precision measurement of the 
$H\gamma \gamma$ coupling would be  an indirect way of 
revealing all the massive 
charged particles that would be present in an extension of the \sm. These 
heavy quanta would not decouple, and would therefore contribute substantially 
to the production rate in \gag, offering a means  for indirectly 
revealing the presence of new heavy particles. 

In $\gamma \gamma$  colliders, the fraction of the initial electron's 
energy that is retained by the 
laser-backscattered photon  can be tuned
by varying the parameters of the lasers. 
{\em If} one assumes that the mass of the Higgs has already been
measured in the \epm\ mode, the photon collider could then be 
precisely designed to sit on the Higgs resonance.
With enough luminosity, one could then conduct 
precision measurements 
of the $H\gamma \gamma$ coupling.
For the IMH,  with the canonical $500$~GeV NLC, this would mean 
operating within a narrow energy range
much below the highest accessible energy (roughly 80\% of the
nominal cms energy of the \epm).
We will refer to this scheme as the narrow-band low-energy  \gag
collider.

Although it is certainly possible to achieve such a peaked \gag\ set-up, 
the question arises whether this is indeed a judicious choice
given that one could have as much as 400~GeV in the $\gam$ cms.
The problem   is that the low-energy narrow band
scheme will preclude   the study of a plethora 
of interesting weak processes\cite{Paris}.
In particular, 
it will not be possible to reach the 
$WW$ threshold (which seems to be a good luminosity monitor) 
and other $W$ reactions that offer a rich physics programme\cite{Paris}.
This could also include 
the direct production of some of those particles that would only 
be probed indirectly 
in $H\gamma \gamma$. Of course, one may argue that these would be necessarily 
produced in the \epm\ mode but in view of the 
known universal character of the production mechanism in \gag,
they may be better studied in \gag. Moreover, it is not excluded 
that the \gag\ mode, when operated in  
the full energy range, can access scalar particles that would kinematically 
be out of 
reach in the \epm\ mode. This could happen if in \epm\ 
they can only be produced in association with another 
heavy  particle. The {\cpviol}-odd Higgs of the minimal supersymmetric 
model is such an example\cite{Gunionzz}.  

It is certain that a 
narrow-band low-energy  \gag\ collider has its merits, 
especially if it is achieved with high luminosity, since precision tests 
on the nature of the light Higgs may be performed. 
Moreover, as we will see, 
with a low-energy scheme many backgrounds are drastically suppressed. 
Investing enough running time in such a mode to  be able to switch 
between different polarization settings (circular/linear polarization,...) one 
could, for instance, {\em directly} test the parity of the Higgs
\cite{Peterparity,Gunionparity} or perform \cpviol tests by probing 
the $H\gam$ coupling\cite{Gunion}.  
These are undoubtedly quite interesting studies to do, but we should 
stress that they do call for very high luminosities and would be done 
at the expense of a rich programme. In addition, keeping in mind that 
this ``narrow-band" scheme presupposes that it is in the \epm\ mode 
that the mass of the Higgs has been determined and used to tune the laser, 
the \epm\  mode would also give  good 
clues on some of the above issues that one wants scrutinized in the peaked 
\gag\ mode. 
For instance, the parity 
of the scalar will, in a large degree, be inferred from its rate of 
production in the \epm\ mode. A spin-$0$ particle, either standard or 
supersymmetric, 
produced through the $VVH$ vertex, is \cpviol even. As pointed out 
in \cite{Gunionzz} for the case of the lightest \cpviol Higgs of the minimal 
supersymmetric model, $h^0$, 
it may also happen that  a measurement of the $h^0\gam$ coupling,
if not very precise, would not provide much more insight. This could occur 
if in the \epm\ mode of a $500$~GeV collider
 only the lightest Higgs of the minimal supersymmetric 
model is discovered while the other SUSY particles are above threshold. 

The choice of the spectrum is then clearly  a critical one  and in some sense
depends on how the \epm\ collider is operating. 
For instance, if the NLC is designed with 
two interaction regions, one devoted to 
\gag\ physics as suggested in \cite{Wiik}, then one should search simultaneously 
in both 
modes or exploit any (expected) earlier evidence in \epm\ to confirm it 
by a selective search in the \gag\ mode. 
One could then 
always dedicate a later (long) run to precision measurements on the Higgs 
properties in a narrow-band low-energy \gag\ set-up. 
Considering the importance of a discovery of the standard Higgs or
of any other scalars such as those that arise in 
supersymmetric models, all means of producing this fundamental
scalar should be explored.
It is therefore essential to address the issue of 
whether 
the intermediate mass Higgs  could be observed as a resonance using
a setting with a photon energy  spectrum that 
allows a whole and self-contained physics programme  to be conducted. 
 A few investigations of this 
aspect have been done with different emphasis and approaches
\cite{Paris,Gunionzz,Borden,Richard,BordenHiggs,Halzen}. 
Here, we reassess the discovery potential of the \sm\ Higgs
at a \gag\ collider
 and give a  complete discussion of
 backgrounds and how one could reduce their effect to a minimum.
We discuss two cms energies for the next linear collider: 
$\sqrt{s_{ee}}=500$ and $350$~GeV\@.
The latter choice could correspond to the energy of a
  top factory and will  
illustrate the numerous advantages of a lower energy machine 
for the IMH search.

While the $\gam$ resonant production of the IMH is straightforward with 
a good cross section, 
problems arise when including the various backgrounds.
Since the IMH will decay predominantly into 
$\bbbar$ pairs, the first obvious background is  direct $\bbbar$ pair 
production, that is, the pure QED process $\gamma\gamma\ra \bbbar$.
This particular background can be easily controlled.
 The solution lies in the observation that the signal receives a 
contribution only  from  $J_Z=0$ (both $\gamma$'s with the same helicity)
while the background comes mainly from
$J_Z=2$. Polarizing the laser so as to obtain a $J_Z=0$ spectrum will
drastically reduce the background. This will be discussed in section 3.
Unfortunately, at photon 
colliders, other large backgrounds occur. These are due   to the 
  hadronic structure  of the photon, meaning that the photon 
 can ``resolve'' into gluons and quarks with
just some spectator jets left over.  Quark pair
production could then occur through $\gamma g$ (so called 1-resolved).
Both photons could also resolve in quarks 
and gluons (2-resolved), introducing many new processes, e.g.
$gg\ra \bbbar, \qqbar \ra \bbbar $.
Although the luminosity for 2-resolved processes is very much suppressed
(see section 3)  the
cross sections involved are proportional to
$\alpha_s^2$ rather than $\alpha^2$ or $\alpha\alpha_s$ for direct
and 1-resolved processes respectively, leading to non-negligible effects.
The 1-resolved background leading to highly boosted
 events   can be controlled
with cuts on a boost related variable. 
Backgrounds from 2-resolved processes
have a distribution that is peaked at small $p_T$ and can
be effectively reduced by a cut on this variable.

Other backgrounds occur in processes
where the $b$ quark is replaced by a charm quark.
One may think, at first, that these 
should not cause any problem since with good $b$-tagging
 the probability that a $c$ quark is misidentified as a $b$ is rather 
low. However, since charm production is much larger than $b$ production,
due to the $c$ quark's stronger coupling to the photon, it remains
an important background.
In this respect the jet tagging strategy used will turn out to be crucial.
The method for pulling a signal out of all this background will be
discussed at length in section 4 with special emphasis on the issues of
polarization of the laser beams, single jet versus double jet $b$-tagging,
efficiencies in $b$-tagging
and resolution in $\bbbar$ invariant mass.
The conclusion that we will reach in section 5 is that, with
a total integrated luminosity of 10~fb$^{-1}$, only
110--140~GeV Higgs could be seen at 500~GeV colliders
while a 350~GeV collider with good parameters could
cover the whole   IMH region, although the case $M_H=90$~GeV calls 
for an optimized set-up of the collider to overcome the background
from the $Z$.


\section{Luminosity spectrum }
 
\renewcommand{\thequation}{\thesection-\arabic{equation}}
\setcounter{equation}{0}

As described in \cite{PhotonCol}, a high-energy photon-photon collider 
can be obtained by  
converting a high-energy electron into a very energetic photon 
through Compton backscattering of an intense laser light.
The luminosity spectrum of the high-energy 
photon depends directly on the differential Compton 
cross section ($\sigma_c$) and is given by
\beqn
f_c(y)&=&\frac{1}{\sigma_c} \frac{{\rm d}{\sigma_c}}{{\rm d}y}=\frac{2}{x_0} 
\frac{\sigma_0}{\sigma_c} C_{00}(y)\;\;\;\mbox{{\rm with}} \; \nonumber \\
C_{00} (y)&=&{1\over 1-y} + 1-y-4r(1-r)-2\;{\bf \lambda_e P_c}\;rx_0(2r-1)(2-y),
\label{c00}
\eeqn
where $\lambda_e$ is the average helicity of the initial electron and $P_c$ is
the degree of circular polarization of the initial laser beam.
The factor 
$\sigma_c/\sigma_0$ takes into account the normalization of the
total spectrum\cite{Paris}. We will assume that each electron is converted
into one photon.
The variable $y$ represents the fraction of the beam energy 
carried by the backscattered photon and 
$$ r = {y\over x_0(1-y)}\leq 1,$$ where $x_0$ is directly related to the 
maximum energy, $\omega_{\max}=x_0/(x_0+1)$, of the ``collider" photon.
In order to reach the highest possible photon 
energies one should aim at having as large a $x_0$ as possible.
The value that we chose, $x_0=4.82$,  is the
highest possible value considering that  for larger $x_0$'s, one 
reaches the threshold for  $e^+e^-$ pair production.
This arises  from interactions between the produced
photon and the laser photon and consequently results in a drastic
drop in the $\gamma\gamma$ luminosity.

The original electron as well as the laser 
can be polarized, resulting in quite distinctive  spectra depending 
on how one chooses the polarizations. 
  One   important observation is that the total energy
spectrum depends on the 
product of the helicities of the electron {\it and} of the photon. 
As a consequence, if either the lasers or the electron beams (which is
more likely) are not polarized at all, the resulting total spectrum will
be the same as if neither were polarized.
However, this does not mean that the  effect of any initial polarization is lost.
Indeed, the backscattered photons will retain 
a certain amount of the polarization of, say, the laser photon beam.
This is because, contrary to the energy spectrum, 
the mean helicity of the produced photon does not depend on the 
{\em product}
of the mean initial helicities. 
Therefore, one can get a dominant helicity configuration for the colliding 
photons by having only the laser polarized, which is very easily obtained. 
The energy-dependent mean helicity is given by 
  $\langle\xi_2\rangle =C_{20}/C_{00}$. 
The function $C_{20}$ writes\cite{PhotonCol,Paris}
 \begin{eqnarray}
 C_{20} = 2\;{\bf \lambda_e}\; rx_0(1+(1-y)(2r-1)^2)
-{\bf P_c} (2r-1)({1\over 1-y}+1-y).
\label{c20}
\end{eqnarray}

 The \gag\ luminosity spectrum is a convolution involving the 
differential Compton cross sections of the two photons  
and is polarization dependent.
 In the case where either the lasers or electron beams are 
unpolarized, one obtains a broad spectrum which is almost a step 
function, that extends almost all the way to the maximum energy (see Fig.~1b). 
  In the case where both arms have 
$2 \lambdae P_c=+1$, the spectrum has a ``bell-like" shape
which favours the middle range values of $\tau=s_{\gam}/s_{ee}$.
With  $2 \lambdae P_c=-1$, one obtains a spectrum that
is peaked at high energies\cite{Paris}.  
For the Higgs search, that is when we would like to keep an almost constant 
value for the differential luminosity, the ``broad" spectrum, 
as that achieved with unpolarized electrons, 
  would be satisfactory.
 However, one should always insist on 
having some polarization since 
polarized laser beams (which are easily obtained) 
and electrons (which should not be too difficult) means that 
the colliding photons are in a preferred state of polarization.
A spectrum  that favours the $J_Z =0$ is highly recommended
since the Higgs is produced in this channel. 

\begin{figure*}[hbt]
\caption{\label{spectre34}{\em {\bf (a)} 
Projecting the contributions of the $J_Z=0$ and the $J_Z=2$ polarized 
spectrum in the ``broad" setting $2\lambda_e P_c=2\lambda_e' P_c'>0$. 
Thick lines are with a $100\%$ longitudinal polarization for the electron 
while the thin lines are for $50\%$ longitudinal polarization. The lasers 
are taken to be fully right-handed. 
{\bf (b)} As in {\bf (a)} but for unpolarized electrons and with  
 a   cut of $|z|<.3$ (see text for the definition of $z$). 
The dot-dashed curve represent also
the spectrum for unpolarized beams.}}
\begin{center}
 \mbox{\epsfxsize=14cm\epsfysize=9.cm\epsffile{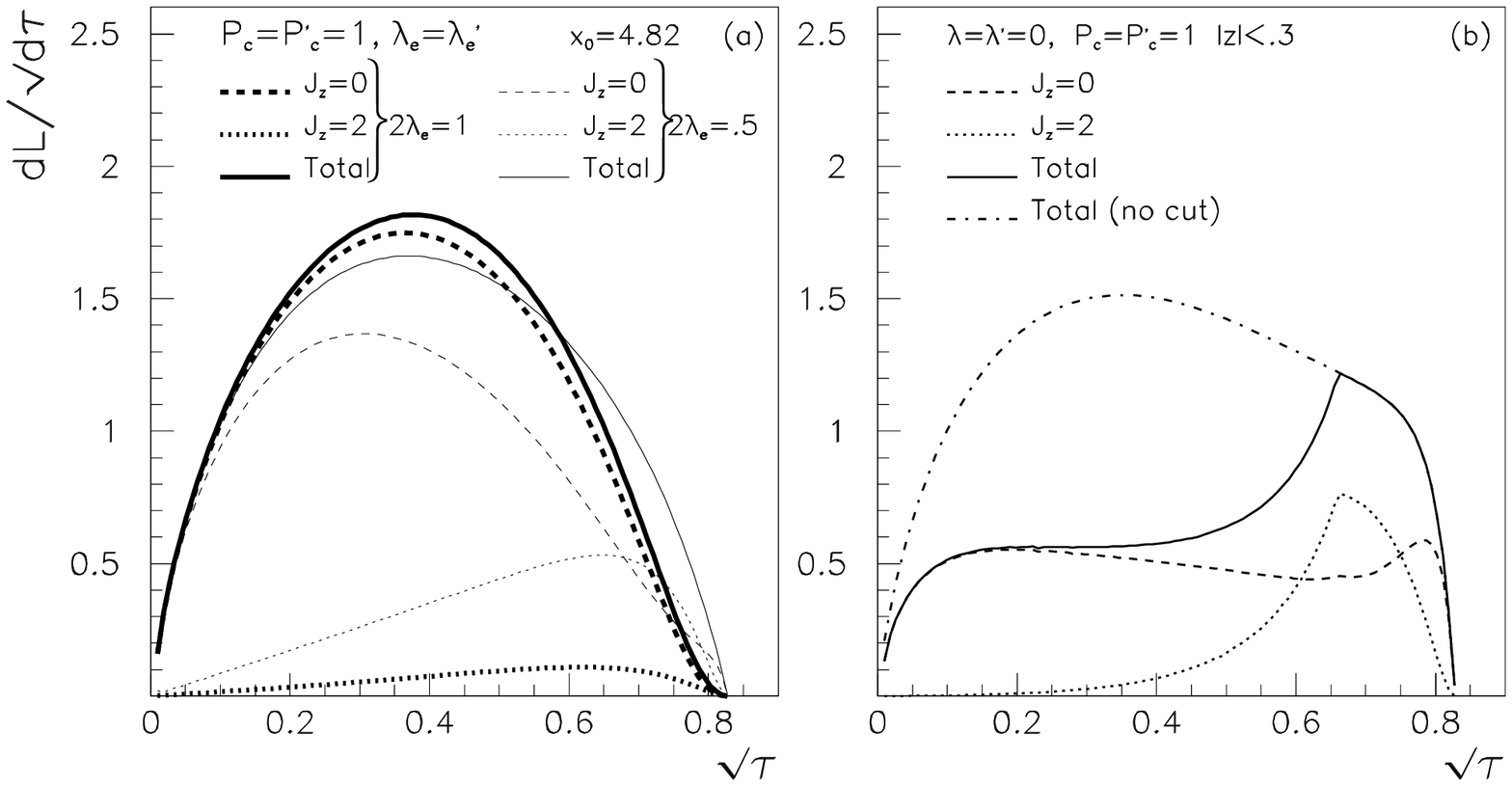}}
 \vspace*{-1.cm}
\end{center}
\end{figure*}

 In the case where both lasers 
are tuned to have a right-handed circular polarization ($P_c=P_c'=+1$), 
 the high-energy photons are produced 
mostly with the same helicity, therefore giving a $J_Z=0$ dominated 
environment. We show in Fig.~\ref{spectre34}a how 
the total luminosity is shared between the two states $J_Z=0$ and $J_Z=2$. 
This near purity of $J_Z=0$ is not much degraded if the 
maximum mean helicity of the electron is not achieved.
     We show on the same 
figure (Fig.~\ref{spectre34}a) 
what happens when we change  both $2\lambda_e$ and
$2\lambda_e'$ from $1$ to $.5$, keeping 
$P_c=P_c'=1$. There
 is still a clear dominance of $J_Z=0$, especially for the lower values 
of the centre-of-mass energy. We would like to draw   attention to the fact 
that this effect (increasing the 
$\frac{J_Z=0}{J_Z=2}$ ratio) 
can be further enhanced (when the maximal electron polarization is not 
available) by imposing cuts on variables that are related
to the boost. The 
point is that the mean helicity of the final photon, $\lambda_\gamma$, 
is non-zero even in the case of 
no electron polarization. Now, if the energy factor 
multiplying $P_c$ in eq.~\ref{c20} is the same for both photons, 
we would expect that 
the colliding photons have the same degree of polarization, hence producing 
a predominantly $J_Z=0$ environment. 
In Fig.~\ref{spectre34}b 
we show the luminosity spectrum for the case 
where the laser photons have the same maximal circular polarization 
while the electrons are unpolarized and where we have imposed a cut 
$|z|<0.3$.
 The variable $z=y_1-y_2$ is directly related to the boost, where
 $y_1$ and $y_2$ are the fractions of the initial beam energy 
retained by each colliding photon.
 We see that for 
small centre-of-mass energies $\sqrt\tau<0.3$, 
we have a highly dominant $J_Z=0$ environment.
This was achieved at the expense of  a  drop in the
luminosity which is mainly due to eliminating the $J_Z=2$ contribution. 
The relevance of this 
observation will be fully exploited in the Higgs search section. 

We conclude this section by a few general remarks on the uncertainties 
introduced by these luminosity functions. First, 
  the spectra we have used are theoretical ones.
It would be extremely important to  verify that the measured spectra do not
deviate
much from the theoretical luminosity calculations. Furthermore, 
a measurement of both the differential (as a function of the 
invariant \gag\ centre-of-mass energy) and total 
\gag\ luminosity as well as
a reconstruction of  the polarized $J_Z=0$ and $J_Z=2$ spectra would be
 useful\cite{Paris}.
A broad spectrum, with a sizeable luminosity
at high energy, would be most favourable for 
the luminosity monitoring. One could then use the  
  $WW$ production  for that purpose.
We have  also  assumed that the density of the laser 
photons is such that all the electrons are converted and that multiple 
scattering is negligible. Furthermore,
 the conversion distance between the interaction
point and the laser hit was taken to be  zero as is   customarily done.
We remark that 
$b$-tagging for instance with a vertex detector is  
a pivotal issue
in the detection of a Higgs with an intermediate mass. We note that this
might be hard to achieve  especially if one needs 
a strong magnetic field very  close to the interaction region
in order to   deflect the initial electrons.


\section{Signal and Background}

\begin{figure*}[htb]
\begin{center}
\caption{\label{Branchingh}{\em 
Some of the \sm\ branching ratios of the Higgs as a function 
of the Higgs mass calculated 
with a top mass of $175$~GeV\@. Also shown (thick lines) are
the total width and the \gag\ width.}} 
  \vspace*{-1.2cm}
 \mbox{\epsfxsize=12.5cm\epsfysize=12cm\epsffile{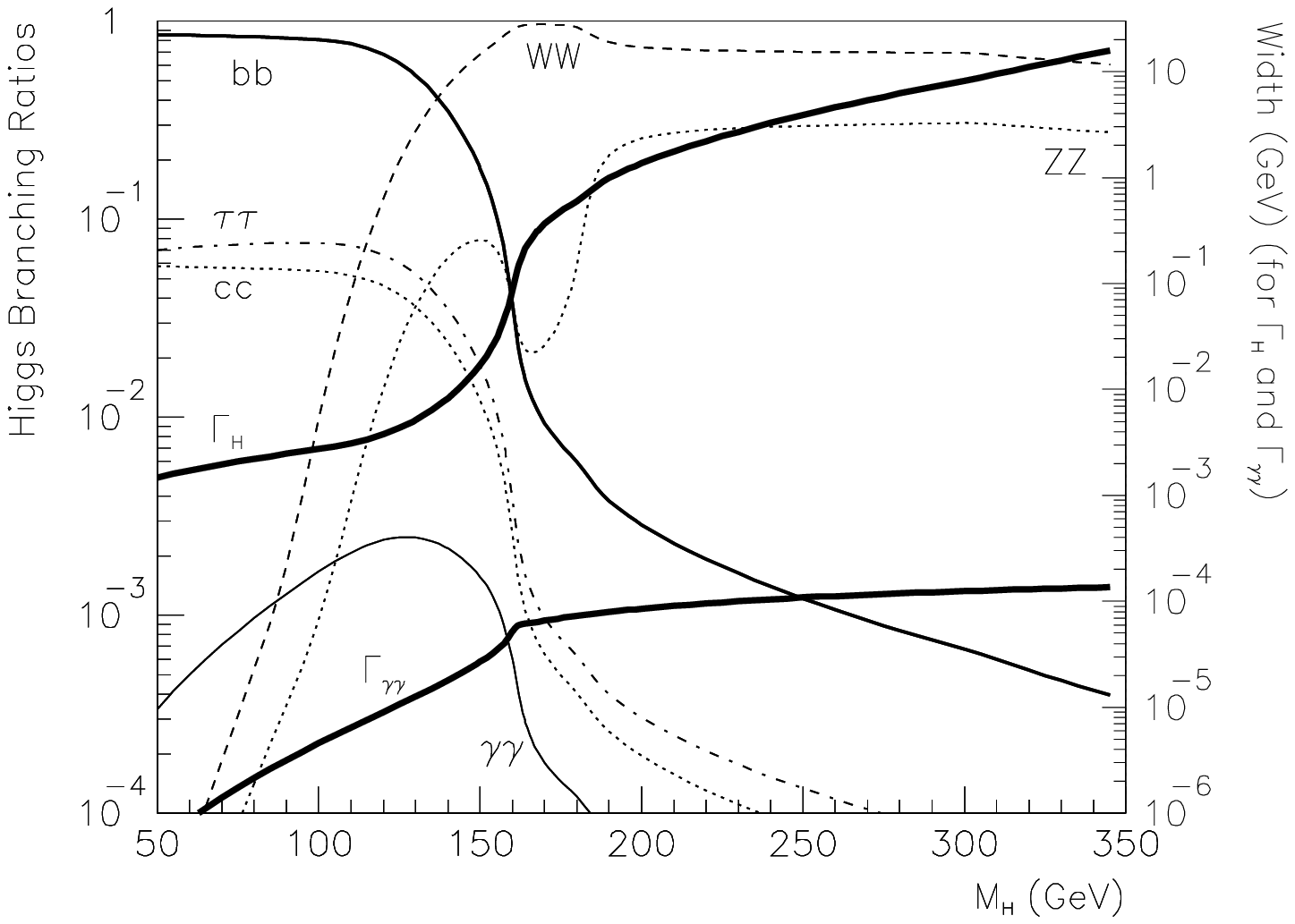}}
 \vspace*{-.5cm}
\end{center}
\end{figure*}

The intermediate mass Higgs will decay predominantly into a 
$b \bar b$ pair and has an extremely narrow width 
(see Fig.~\ref{Branchingh}). This width, 
$\Gamma_H=\Gamma_{\mbox{\rm \scriptsize total}}$, is of order of a few MeV\@. 
In the rest frame of the Higgs, the fermions are produced isotropically in 
the $J_Z=0$ state. The corresponding cross section is described by 
the Breit-Wigner formula\cite{Gunionzz,BordenHiggs}
\beqn
\sigma{(\gamma \gamma \stackrel{H}{\longrightarrow} b \bar b})=
\frac{8 \pi \Gamma(H\ra \gam) \Gamma(H\ra b\bar{b})}{(\she -\Mhe)^2+
\Gamma_H^2 \Mhe^2} (1+\lambda_1 \lambda_2).
\eeqn
The exact expressions for the branching ratios of the Higgs and the 
$\Gamma(H\ra \gam)$ width including QCD corrections 
that we will use in our analysis 
are taken from\cite{Abdel}\footnote{We thank Abdel 
Djouadi for providing us with the Fortran code.}. A top mass of $175$~GeV 
has been assumed. A change to $m_t=150$~GeV hardly affects our results. 
 
As the width of the Higgs, $\Gamma_H$, is very small,  
to simulate the finite resolution in the $b\bar{b}$ invariant mass 
as it would occur in an experimental set-up,
we have introduced a Gaussian smearing of the Higgs signal.
The cross section, $\tilde{\sigma}$, as a function of 
  the invariant mass of the $b\bar{b}$ pair,  $M_{b\bar{b}}$ ,
is  calculated using
\beqn
\frac{d\tilde{\sigma}}{d M_{b\bar{b}}}=\int
d\sqrt{\hat{s}}\frac{1}{\sqrt{2\pi}\delta} \exp\left[
-\frac{(M_{b\bar{b}}-\sqrt{\hat{s}})^2}{2\delta^2}\right] \;\;
 \frac{d\sigma}{d\sqrt{\hat{s}}},
\eeqn
where
$\delta$ is the detector resolution, and $\sigma$ is given  in 3.3. 
 We consider two 
values for the resolution: $\delta =5$~GeV and $\delta =10$~GeV\@.
The first number is comparable with  what can now be achieved at LEP
with a microvertex detector
\footnote{We thank Marie-No\"elle Minard for clarification on this
issue: the point is that $b$-tagging provides kinematical constraints that
improve the $m_{b\bar{b}}$ resolution.}
while the second one is more conservative.
It will turn out than our conclusions depend critically on the would-be achieved resolution.

Note that the IMH could also decay in $\ccbar$ pairs  but this is suppressed
by an order of magnitude relative 
to $\bbbar$. Moreover, there is an overwhelming $c\bar{c}$ continuum that has 
to be drastically reduced. This background suppression makes the 
 $\ccbar$  contribution from the Higgs totally negligible.  


The cross section for Higgs production at $500$~GeV ($350$~GeV) 
with a polarized spectrum ($2\lambda_e=.9, P_c=1$) is roughly 
$\approx 35$~fb (50~fb)
for $m_H=120$~GeV and decreases slowly  for other values
of the IMH. 
With a luminosity of 10~fb$^{-1}$ a large sample of scalars 
should be collected. 
The problem is how to extract the signal from the background.

\begin{figure*}[htb]
\begin{center}
\caption{\label{ggh1}{\em The Higgs signal into $b \bar b$ and its QED background}} 
\vspace*{.2cm}
\mbox{\epsfxsize=11.cm\epsfysize=3cm\epsffile{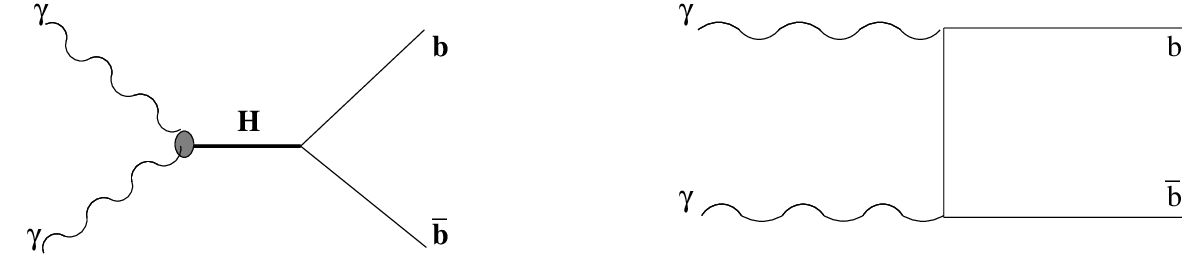}}
\vspace*{-.5cm}
\end{center}
\end{figure*}
 
\subsection{Background} 
The most obvious background is the direct QED quark pair production
 $\gam \ra b \bar b$ (Fig.~\ref{ggh1}).
A glance at the corresponding expression for the differential cross section 
gives a clue 
as to how one could efficiently suppress this background. For the quark 
of charge 
$e_f$ and with $N_c=3$ we have, in the \gag\ {\em cms} with 
$\theta^*$ being 
the $q$ scattering angle and $x=\cos\theta^*$,
\begin{eqnarray} 
\frac{{\rm d}\sigma_{QED}}{{\rm d}x}=\frac{2 \pi \alpha^2 e_f^4 N_c \beta}
{\she (1-\beta^2 x^2)^2} \left\{(1+\la_1 \la_2)(1-\beta^4)+
(1-\la_1 \la_2)\beta^2 (1-x^2)(2-\beta^2 (1-x^2))\right\}.\nonumber\\
\label{ggbbqed}
\end{eqnarray} 
It is clear that the bulk of the cross section is from the 
extreme forward-backward region. A modest cut on $\cos\theta^*$ 
will reduce the continuum substantially and will almost 
totally eliminate its 
$J_Z=0$ contribution (note the $(1-\beta^4)$ chiral factor). 
Therefore, 
choosing a spectrum with a predominant $J_Z=0$ component\cite{Gunionzz} and
applying a cut on  $\cos\theta^*$, or alternatively on $p_T$,
 should do the trick. It is instructive 
to note that the $J_Z=2$ contribution, because of angular momentum conservation, 
vanishes in the exact forward region. 
 One should also worry about  production of light quarks if no flavour
identification is possible.  In fact, even in the case where $b$-tagging
is available, since it can never be perfect, the 
charm quark causes much problem, especially that its rate of production (direct)
is roughly 
16 times larger than $b$ pair production.
We will see in the following that this will in many cases be a  major
background.

\begin{figure*}[htb]
\begin{center}
\vspace*{-.5cm}
\caption{\label{ggh2}{\em Bottom pair production through  
``1-resolved"  photon.}}
\vspace*{.2cm}
\mbox{\epsfxsize=13.cm\epsfysize=2.5cm\epsffile{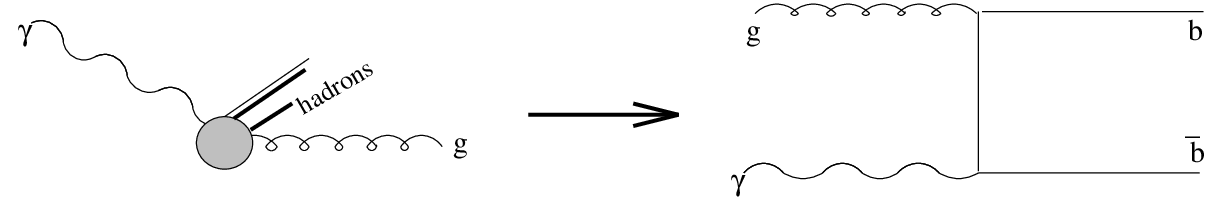}}
\vspace*{-.5cm}
\end{center}
\end{figure*}

It has recently been pointed out\cite{Halzen} 
that, unfortunately, this is not the whole 
story. Owing to the fact that the photon has a hadronic 
structure\cite{Witten} 
 it can 
``resolve" into a gluon or quark with some spectator jets   
left over. One then has to take into account 
processes like $q \bar q$ production through $\gamma g$ (see Fig.~\ref{ggh2}),
as well as a host of 1-resolved and 2-resolved processes listed
in Table 1. 
In an obvious notation, 2-resolved refers to processes where
both photons resolve into quarks or gluons.
In principle, one could discriminate the gluon or quark initiated
 processes from 
the direct ones through the presence of the spectator jets. However, 
it seems to be very difficult to tag these extreme forward spectator jets in the \gag\ 
environment\footnote{For a more optimistic view, see \cite{BordenHiggs}.}.

 \begin{table*}[htb]
\caption{\label{backgroundlist}
 {\em  Background to  the Higgs signal. 
  The corresponding processes with antiquarks replacing
quarks are understood and q=u,d,s,c,b.}}
\begin{center} 
\vspace*{0.3cm}
\begin{tabular}{|c|c|c|}\hline 
 &$b$ quarks  &$c$ quarks\\\hline
& &\\ 
Direct& $\gam\ra\bbbar$ & $\gam\ra\ccbar$\\\hline
& &\\
1-Resolved & $\gamma g\ra\bbbar$ & $\gamma g\ra\ccbar$\\
 & $\gamma b\ra  g b$ & $\gamma c\ra g c $\\\hline
& &\\
 & $g g\ra\bbbar$ & $g g\ra\ccbar$\\ 
 & $\qqbar \ra\bbbar$ & $\qqbar \ra\ccbar$\\   
 & $ bb \ra bb $ & $ cc \ra cc $\\
  
 2-Resolved & $g  b \ra g  b  $ & $g  c \ra g c  $   \\
 & $q  b \ra q b  $ & $q  c \ra q c  $   \\
 & $Z\ra \bbbar  $ &  $Z \ra \ccbar$  \\
 &  $(W\ra c\bar{b}) $ & $W \ra c\bar{s} $ \\\hline 
\end{tabular}
\end{center}
\end{table*}

\begin{figure*}[htb]
   \begin{center}
   \vspace*{-.5cm}
  \caption{\label{lumresolved}  {\em  (a)
Luminosities for 1-resolved processes after integration
 over the unpolarized 
photon spectrum. Thick lines are for processes with
gluons, full lines for u quarks, dots for c quarks, dash
for b and dash-dot for d }.  
{\em (b) Luminosities for a selection of 2-resolved processes,
same labelling as in (a), dash-dot is for uc. }}
 
 \vspace*{-1.cm}
\hspace*{-1.2cm}
\mbox{   
\mbox{\epsfig{file=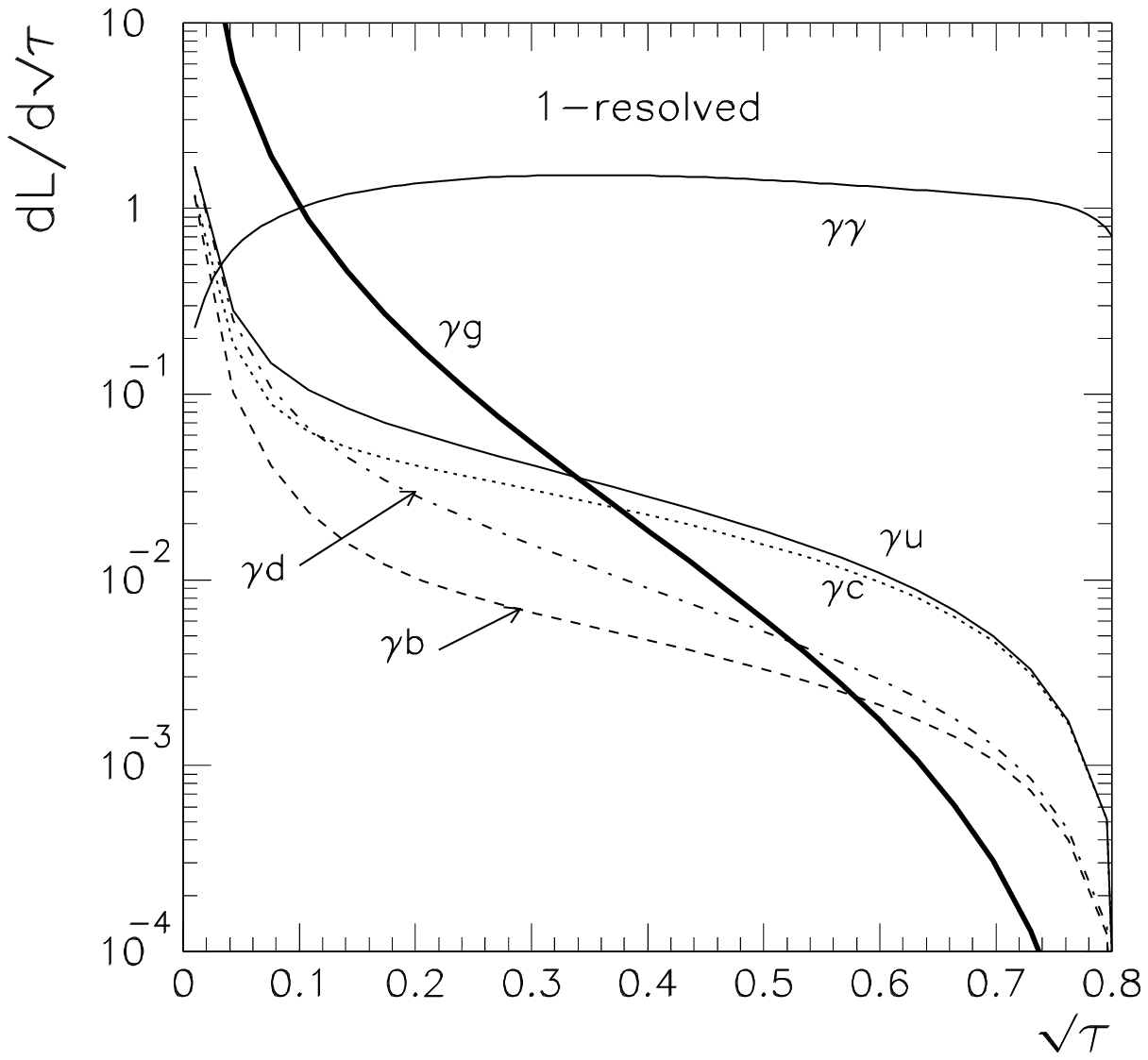,width=9.5cm,height=11cm}}
\hspace*{-1.5cm}
\mbox{\epsfig{file=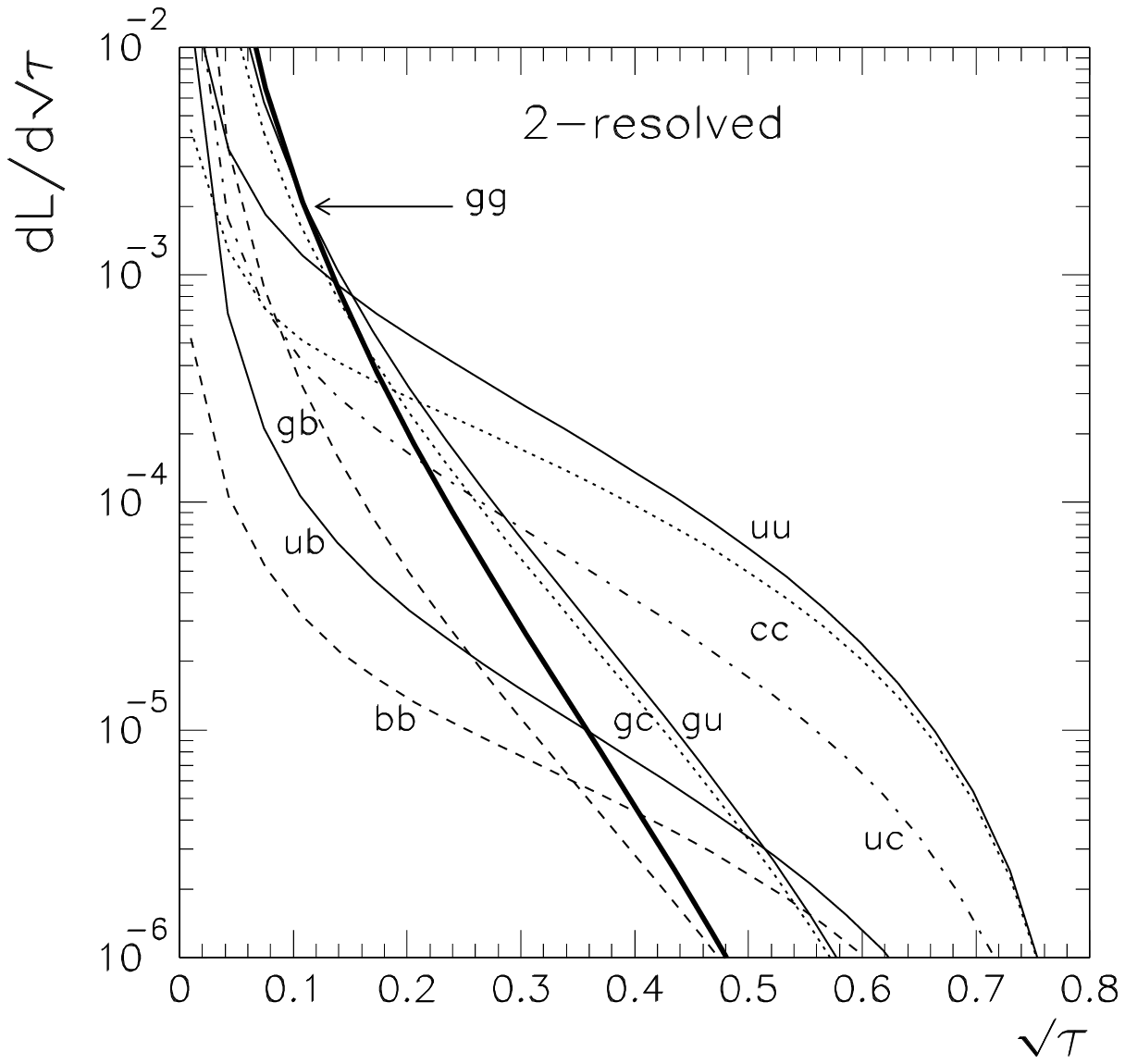,width=9.5cm,height=11cm}}}
   \end{center}
   \vspace*{-1.cm}
\end{figure*}

To get an idea of the relative importance of the various resolved contribution
we first show in Fig.~\ref{lumresolved} the effective
luminosities for all 1-resolved and a sample
of the more relevant 2-resolved processes. These should be compared with a
$\gam$ luminosity of order one, also shown in Fig.~\ref{lumresolved}.  
These curves were obtained   using the GRV distribution functions\cite{GRV} 
with $Q^2=(60$~GeV)$^2$ where the same
set is taken for quarks and antiquarks.  The choice of $Q^2=(60$~GeV$)^2$
corresponds to the average $M_H/2$ of Higgs masses that we are covering.
We have found that the influence of using another set of structure
functions\cite{AFG} as well as varying the value of $Q^2$
is minimal (see below).
The photon luminosities with unpolarized photons were folded in. 
To lowest order the structure functions are directly related
to the charge. One therefore expects the $u,c$ content of the photon
to exceed the $d,s,b$ content.
The luminosity would not change much even  with  maximum polarization
($2\lambda_e=0.9$ and $\lambda_\gamma=1$) since the polarized
photon energy  spectrum does not  differ significantly from the unpolarized one.
As always, we chose
$x_0=4.82$ so that $\sqrt{\tau_{max}}=\sqrt{s_{\gam}^{max}/s_{ee}}\sim 
0.83$.

The first thing to note here is 
the important luminosity of $\gamma g$ at small $\sqrt{\tau}$
and the fast drop of the  $\gamma g$ luminosity with 
energy. In contrast, the $\gamma q$ luminosities, 
which are smaller at lower $\sqrt{\tau}$, decrease  more gradually.
Although the photon transfers only a small fraction, 
$y_g$, of its energy to the gluon, at $\sqrt{s_{ee}}\sim 500$~GeV 
the overall energy can still be 
large enough for this gluon to combine with a photon 
leading to a substantial luminosity at the
subsystem energy, that corresponds to the IMH production.
Quark pair production initiated by $\gamma g$ would
then constitute an important background.
From these graphs it is also  clear that it will be much easier
to pull out a signal for the IMH in a lower energy
collider (here at $350$~GeV) as the resolved contribution,
for the same subsystem energy
will be suppressed  compared with the $500$~GeV version. 
For 2-resolved processes, the decrease of the luminosity
with energy is even more drastic. Although, in the
region of interest, the luminosities for 2-resolved 
processes containing $u,c$ quarks or gluons are suppressed
by more then two orders of magnitude relative
to 1-resolved processes (say $\gamma u$), they will
contribute significantly to the background, 
especially at $500$~GeV\@.

With a large effective luminosity for quark pair production, one might
think that it would even be possible to also produce a Higgs   through
$\qqbar$ fusion. The coupling $H\ra \ccbar$, for example, is several 
orders of
magnitude larger than $H\ra \gam$. Nevertheless, the luminosity is just too
small to have a meaningful production rate. We 
 found that the cross section
for Higgs production from $\ccbar$ would be a mere 1\% of the
$\gam$ process and less than 0.1\%  for $\bbbar$.

\begin{figure*}
   \begin{center}
   \vspace*{-1.cm}
  \caption{\label{backg500}   {\em 
Signal and backgrounds for $H\ra \bbbar$  at $500$~GeV with polarized
photon spectrum, $M_H=120$~GeV and $\delta=5$~GeV\@.
Backgrounds include direct (dot), 1-resolved (dash and dot-dash)
as well as 2-resolved (full) processes. $Z\ra \bbbar(\ccbar)$
with $\delta=5$~GeV is also shown.  Note that for the Higgs signal, we
have cut the tail due to the smearing so that all the events fall within
$\pm 2\delta$.}}
 \mbox{\epsfxsize=14cm\epsfysize=18cm\epsffile{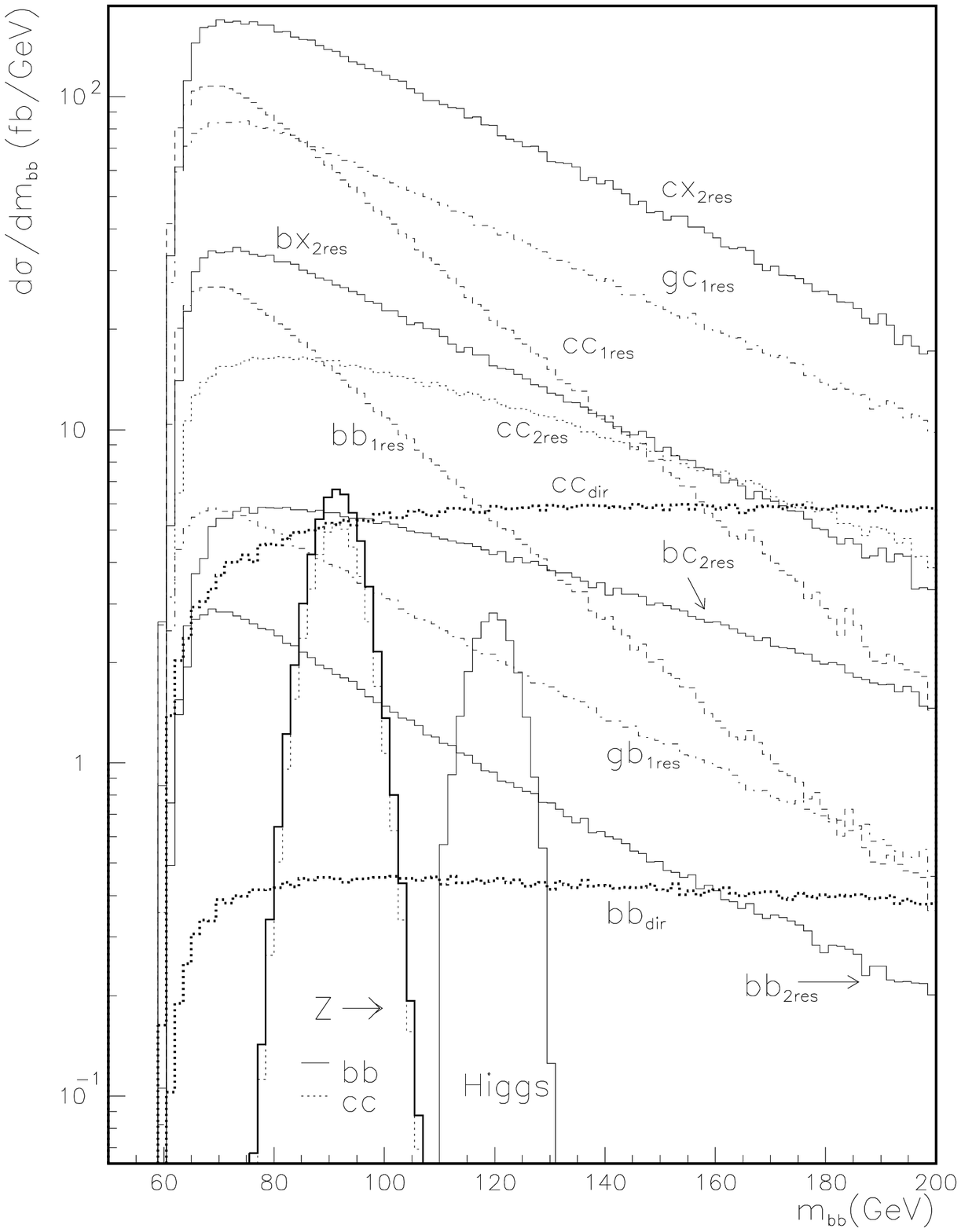}}
 \vspace*{-2cm}
   \end{center}
\end{figure*}

The cross sections for all backgrounds at $500$~GeV are shown in 
Fig.~\ref{backg500} before applying any cuts except for
 a cut on $p_T<30$~GeV to avoid 
the $t$-channel singularity that may occur in the forward region.
 Here the luminosity functions have already been
folded in and we have made use of the polarization of the initial beams
($\lambda_e=0.9, P_c=1$).
 The only significant changes in the cross sections
with unpolarized beams would be   roughly 
an order of magnitude enhancement of the direct processes.
  The cross sections for 2-resolved processes
implicitly include a sum over all initial quarks, and only
the QCD processes at tree-level, which contribute by far the most to the
cross sections, were calculated, in order to simplify the
computation. To be consistent with the calculation of the subprocesses
done at tree-level, we only included   the leading log
contribution to the structure functions.
 
 A glance at the figures suffices to realize
the formidable task that we are facing in order to extract the Higgs signal.
A closer look at Fig.~\ref{backg500} suggests an immediate
way to suppress the background: one should require a double-jet tagging strategy.
Double-jet tagging means that in order to keep an event, {\em both} jets
must be identified as $b$.  With this method,
all final states with a light quark or gluon are 
rejected with a high efficiency. 
If one could achieve perfect rejection of the light quarks and gluons, the
 remaining background would be $q \bar{q'}$ ($q,q'=b,c$)  production 
 from  $\gamma g$, 2-resolved and direct QED,
in order of decreasing importance.  
The relative contribution of $b$ or $c$ jets depends on
the tagging efficiencies used.
Note that polarized spectra have already been taken into 
account to reduce the direct contribution.
 Except for direct quark production, all backgrounds
are more severe for lower invariant masses. We
therefore expect the lighter Higgs to be
much harder to see.
Obviously the 2-resolved processes cannot be rejected off-hand
as they constitute the dominant background.
 In particular, the  largest cross section at all energies is from
2-resolved $cx$, where $x$ stands for any light quark or gluon.
In fact this cross section is largely dominated by $g c\ra g c$.
We stress that this is so only before 
neither $b$-tagging nor cuts are applied.
The relative importance of the
2-resolved contribution is due to  the   very large
cross section of some of the subprocesses. Since they
involve only quarks and gluons, they are enhanced by a factor
$(\alpha_{s}/\alpha)^2$ over the direct processes.
Furthermore, since a  sum over initial quarks 
must  be performed, many subprocesses contribute here.
The ones initiated by gluons, $u$ or $c$ quarks in particular,
are the most important. 
 Luckily, the situation is not as bad as it looks since the main 
2-resolved processes are those with a $t$-channel which contribute
for the most part
 in the small $p_T$ region. A cut on $p_T$ is an effective
way to reject this type of background.

If the photon transferred   all of its polarization to the gluon
then there would  not   be much 
problem in eliminating the background $\gamma g \ra q\bar{q}$,
 as we will be in the same situation as with the polarized 
\gag\ initiated 
process. Unfortunately, we expect the polarization to be diluted in 
the transfer.  
On the other hand, as 
remarked in  \cite{Halzen}, in the $\gamma g$ initiated process, the gluon 
has in general much less energy than the photon, since the gluon 
distribution inside the photon comes essentially from the low $y_g$ 
region. This will lead to a larger boost of the $q \bar q$ 
system along the photon direction giving a system with 
a much larger rapidity than in the direct processes. Thus the 
authors of \cite{Halzen} have suggested to apply a cut to reject $b$'s with 
large rapidities. We have preferred to use another variable that
is more directly related to the boost, that is, $p_Z$, 
where $p_Z=|p_{Z1}|-|p_{Z2}|$ and
 $p_{Zi}$ are the longitudinal momentum of each jet.
 For other 1-resolved backgrounds the $p_Z$ 
cut would not be as effective since the
resolved quarks  are less boosted than the gluons.
A cut on $p_Z$ or any other variable related to the boost will 
have another beneficial effect: rejecting the direct
$\bbbar$ process. As we have shown in
section 2, this is because such a cut improves on the rejection of the
$J_Z=2$ which dominates the direct quark pair production.
Note that by using the same argument,  if the gluon
retained a fraction of the photon polarization, a 
cut on highly boosted events
would also  further improve  the background rejection 
of the $\gamma g$ initiated process.
 In that sense
the following analysis which assumes
no polarization transfer  is a worst-case scenario.

Before going into more details about the background suppression
methods and the cuts used we should mention that when
 the mass of the Higgs is around that of the $Z$ 
we have another non-negligible background\cite{Ilyazbb}: $Z$ radiation 
off a fermion pair while the external fermions go down the beam undetected.
 The $Z$ subsequently decays into $b \bar b$ or $\ccbar$.
In the case where the fermion is a lepton, 
we have computed the polarized $\gamma\gamma\ra Z l\bar{l}$ cross section
exactly, keeping the full spin information to be able to include
the $Z$ decay into $b\bar{b}$.  The helicity elements were
produced through Madgraph\cite{madgraph}.  All events
where the leptons were less than $10$ degrees from the beam
were included in the background.
The results,  for the unpolarized case, are in 
agreement with the approximate calculation of the same 
cross section based on the splitting functions that describe 
the lepton content in the photon\cite{Ilyazbb}.  
In the case where the associated fermion is a quark, 
we could not   
restrict  ourselves to a tree level perturbative calculation 
or to the use of a naive first order splitting function as for
the leptons.  Both  methods underestimate the
quark content of the photon, especially in the region
of low energy transfer. Rather, 
the quark contributions were calculated  
 using the GRV structure functions (with $Q^2=(60$~GeV)$^2$).
We found that $Zee$ constitute  
$\approx 75\%$ of the total $Z$ cross section for the two energies of interest.
This is consistent with what was found in \cite{Ilyazbb}
at lower energies. The importance of
the electron contribution  is related to the fact that the
$t$-channel peaking in the forward region is more pronounced.
The cross sections summed over initial fermions are shown in
Fig.~\ref{backg500}.  
To simulate realistic resolutions we have once
again smeared the $Z$ signal with  a Gaussian, taking the same
resolutions as for the  
Higgs.{\footnote{We have neglected the width of the $Z$ in doing so, 
since the width is smaller than the resolution expected.}}
 The $Z$ is a  very important background
that is  furthermore hard to cut since 
the difference in the angular distribution, ($1+\cos^2\theta^*$) for
 the transverse
$Z$ and isotropic for the Higgs, is not important enough to eliminate
a significant amount of $Z$ events.{\footnote 
{This also shows that neutral gauge bosons can be easily produced
with a large cross section in $\gam$ colliders. 
A complete discussion of neutral bosons production
($Z,Z'$) will be given in\cite{VincentZ}.}} 
 Fortunately,  with a good resolution for both the $Z$ and the Higgs
this background is relevant only when the two masses are within $20$~GeV\@.

There is another background that is not shown in Fig.~\ref{backg500}:
$W$ radiation off a fermion pair. This background is in some ways
similar to that of the $Z$ but is 
much less important  for three reasons. First, 
the production cross section is smaller since it can occur 
only through quarks, rather than leptons (since the photon
does not couple to the neutrino). 
 Second,  the mass of the $W$ is  $80$~GeV so 
 there is little overlap with
the Higgs for the mass range of interest. Finally the
signature $W\ra c\bar{s}$ will be eliminated almost completely
 by the double tagging strategy while $W\ra c\bar{b}$ is suppressed by a small
quark mixing angle.

\begin{figure*}
   \begin{center}
   \vspace*{-.5cm}
  \caption{\label{backg350} {\em 
As in Fig.~\protect\ref{backg500} but for the case of a 350~GeV \epm collider.}}
 \mbox{\epsfxsize=14cm\epsfysize=20cm\epsffile{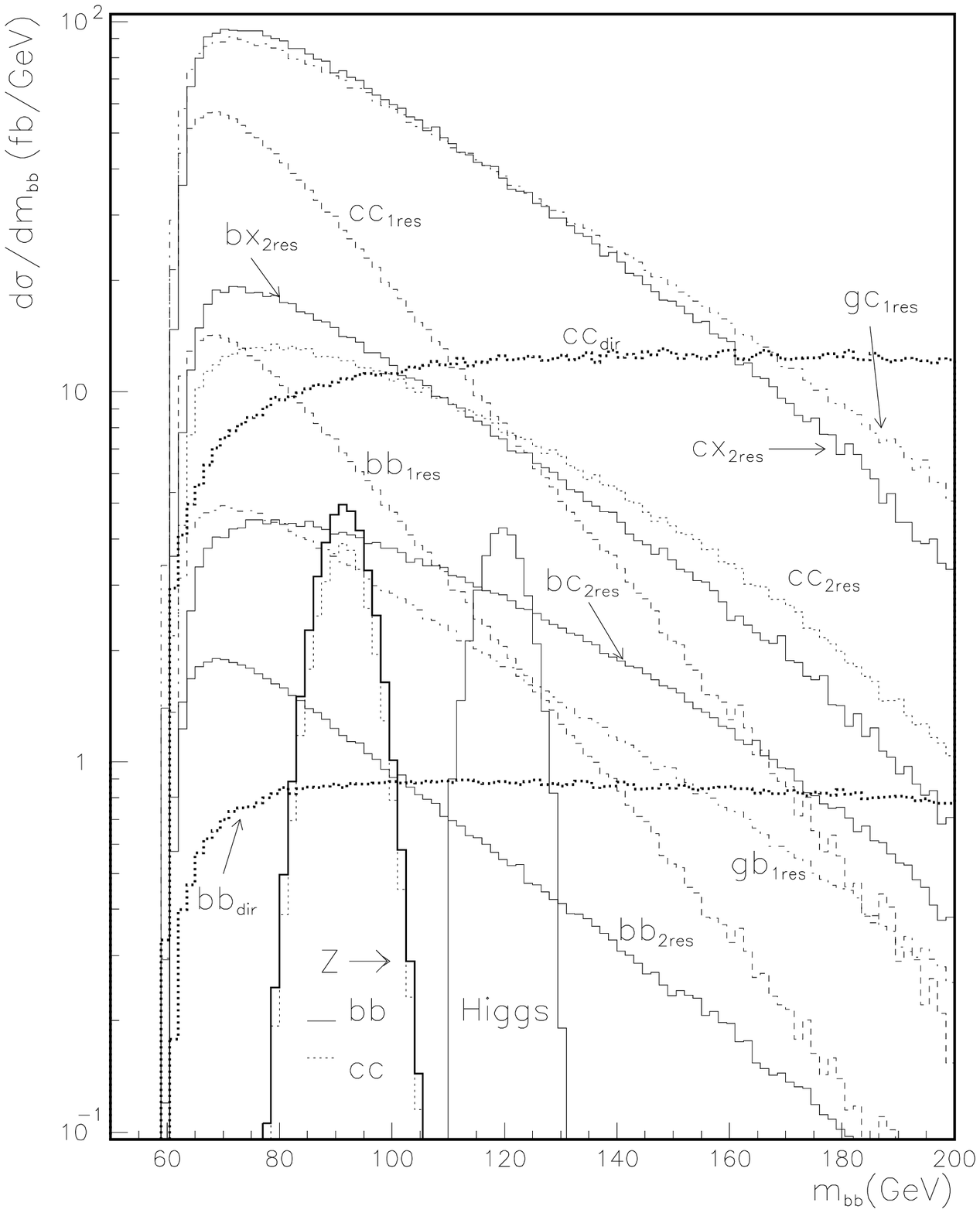}}
\vspace*{-1.2cm}
   \end{center}
\end{figure*}

The previous discussion on backgrounds corresponds to the higher energy version
of
the \epm\ collider. With a lower energy machine the picture differs
 appreciably. For a  lower center of mass energy
in \epm, the same invariant mass  will correspond to a
higher $\sqrt\tau$, hence a lower luminosity 
  for resolved quarks and especially gluons.
 Therefore,  one expects all resolved
processes to be reduced as can be seen in Fig.~\ref{backg350}.
 Here again, a double-tag strategy and an excellent efficiency for rejecting
light quarks will eliminate all heavy+light quarks or gluon
backgrounds. With a perfect rejection of the light quark background, as for
the higher energy machine, the
main background remains $\qqbar$ production. However,
the relative contribution of direct and resolved processes
differs. At $350$~GeV, the
 direct process, which at low invariant
mass has the smallest cross section, takes over the 2-resolved one
for $M_{\bbbar}>110$~GeV and the 1-resolved one for $M_{\bbbar}>140$~GeV\@.
This is so even after polarization has been  used to control the
direct process. 
 As before, a  $p_Z$ cut will dampen both the 1-resolved and direct
$\bbbar$ backgrounds while a $p_T$ cut will reduce the
2-resolved one to a negligible level.
 At the end of the day we expect a much better signal/background
ratio for a lower energy machine.

\subsection{Uncertainties in estimating the backgrounds}

\begin{figure*}[htb]
\begin{center}
\caption{\label{GRVAFG}{\em 
Variation of the a) $\gamma g\ra \bbbar$, b) $cc\ra cc$
cross sections with the choice of structure functions
and value of $Q^2$. The two curves show the
maximum deviation, the full line is for GRV with
$Q^2=(60$~GeV$)^2$, the dotted line for AFG with
$Q^2=(200$~GeV$)^2$.}} 
  \vspace*{-1.2cm}
 \mbox{\epsfxsize=16cm\epsfysize=11.5cm\epsffile{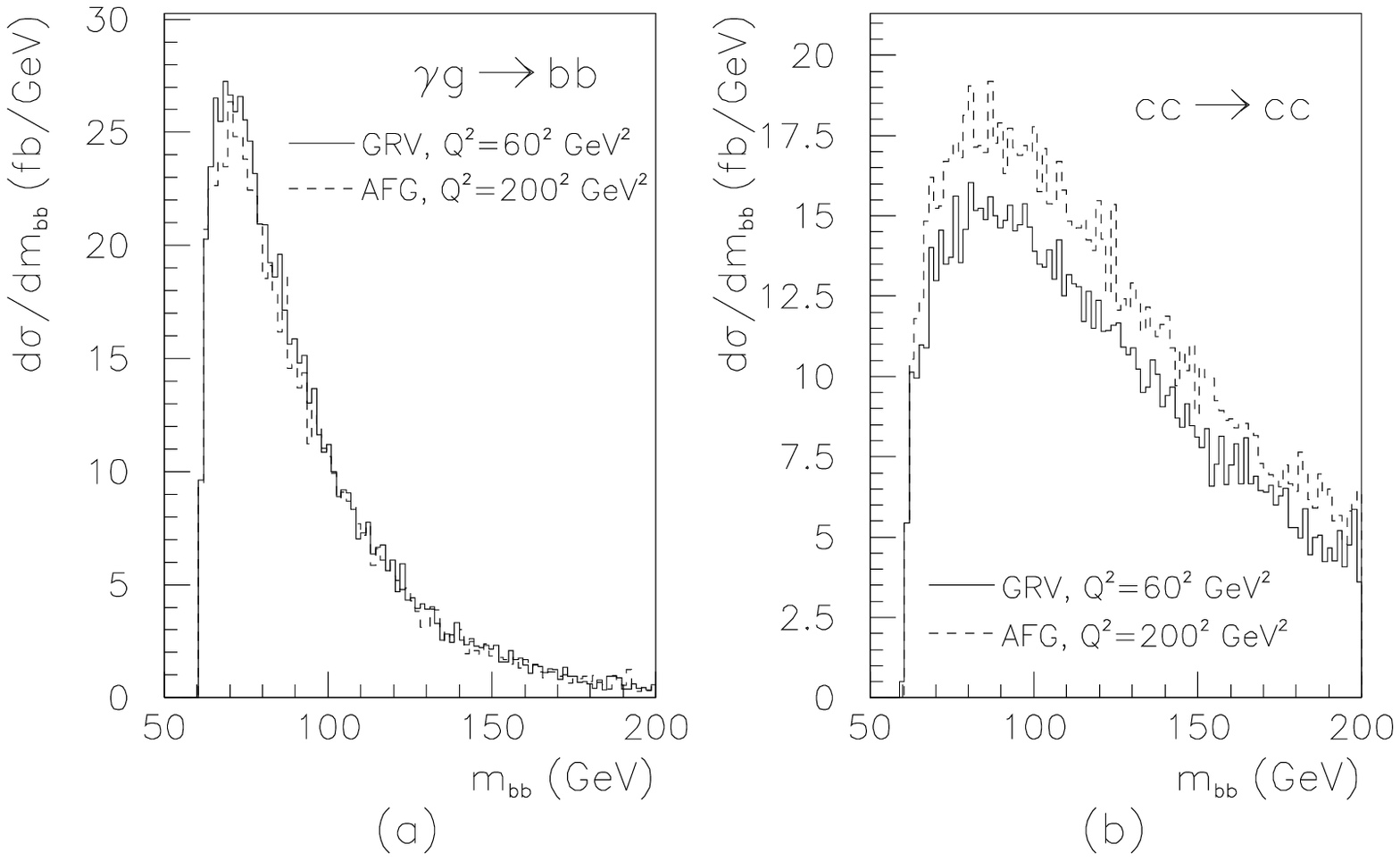}}
\end{center}
\end{figure*}

As mentioned, all the curves for the luminosities and  cross sections
were obtained by taking $Q^2=(60$~GeV)$^2$ corresponding roughly to
the central value of the range considered for $M_H/2$.
We have checked that our conclusions were insensitive to both
the choice of structure functions and
the choice of $Q^2$, within the region $Q^2>p_T^2$. 
For example, the $\gamma g \ra \bbbar$ differential cross section,
which constitutes one of the main backgrounds,  varies
by less than $5\%$ in the $M_{\bbbar}$ 
region of interest when  $Q^2$ is varied
from $(60$~GeV$)^2$ all the way to ($200$~GeV$)^2$ with either the GRV
\cite{GRV} or AFG\cite{AFG} set of
structure functions (see Fig.~\ref{GRVAFG}a). 
The reason is that the $\log(Q^2)$ dependence in the first order
structure function is compensated by the same factor in the
running of $\alpha_s$. It would only be
for very small $Q^2$, where the non-perturbative part
is more important, that we would expect a $Q^2$ dependence. 
This region is irrelevant for the present analysis.
It is the structure function for $c$ quarks that
depends more strongly on the choice of parametrization\cite{AFG}.
In Fig.~\ref{GRVAFG}b, two curves
illustrate the  maximum variation of the cross 
section  for $cc\ra cc$.  The lower prediction 
corresponds to GRV with $Q^2=(60$~GeV$)^2$ while the
higher one to AFG with $Q^2=(200$~GeV$)^2$.
The difference between the two extreme cases is around
$15\%$ for the range considered.
However, since after cuts the 2-resolved background will be 
significantly smaller than
the 1-resolved one, the global effect on the background will be
 very small. We estimate that the total uncertainty
in our analysis never exceeds a few percent.

In  our discussion of background we have not included
QCD corrections to $\gam \ra \qqbar$, either loop corrections or
terms where a gluon is emitted collinear to the quark jet.
This has very recently been studied for the case of
a lower energy photon collider operating at around the
Higgs resonance\cite{jikiaqcd,BordenKhoze}, having in mind 
the measurement of the Higgs couplings where precision is
critical.
It is found that the $J_Z=0$ cross section, that is very small
at tree-level, receives a very large enhancement
(for $c$ quarks more than an order of magnitude in
the region of interest) 
while the correction to the dominant $J_Z=2$ contribution is
very small.
We have estimated the effect of
  neglecting the QCD corrections by taking the most severe variation
of the cross section and assuming that the corrections did not modify
the   angular distribution of the direct process.
Even an increase by a factor of 70 of the direct $J_Z=0$ $\ccbar$ cross section
and a factor of 3 for the $\bbbar$ one, did not
affect our results much. 
For example, the significance $S/\sqrt{B}$   changes 
from 4.0 (4.5) to 3.9 (4.3) for
$m_H=120 (140)$~GeV and $\delta=5$ at 500~GeV (all results for the
significance will be presented in section 5). 
At 350~GeV, and for the same masses, the variation is from
8.1 (8.5) to 7.6 (8.1).
The little influence of such a large enhancement can be
understood rather easily since the major contribution
to the background is the 1-resolved, not the direct. 
Furthermore, even with the large enhancement
and after folding the luminosity spectrum 
the $J_Z=0$ becomes 
only as large as the $J_Z=2$ ({\it i.e.,\ }the net effect is at most 
a doubling of the direct contribution).
Due to the more important role of the direct process at
lower energies we indeed expect the tree-level approximation
to be worse at lower energies. 
One could also be concerned about the QCD  corrections to the $\gamma g$
cross section, which are expected to be as important.
However, since the laser polarization is not retained by the gluon, for this
case the main contribution would essentially be given by the tree level $J_Z=2$.

\section{Background suppression}
In the preceding section we have alluded to some of the means at our 
disposal for 
eliminating various backgrounds; here we develop our full strategy.
The issue of polarization has already been discussed:
a perfect polarization doubles the signal while 
suppressing the direct QED background.
We will consider $P_c=P_c'=1$, which can be achieved easily, and 
$2\lambda_e P_c=2\lambda_e' P_c' \ge 0$ with
$2\lambda_e$ varying from 0 to 0.9. This will be compared 
with the
case of no polarization at all, $P_c=P_c'=\lambda_e =\lambda_e'=0$.
The issue of double {\it vs} single-jet tag, $b$-tagging efficiencies
 and cuts were all investigated.

\subsection{Tagging}

In the discussion of background and its suppression, one key question
is the tag strategy. Single $b$-tagging, for which
one requires that only one jet is identified as a $b$,
implies that one can  keep  both a  larger 
proportion of the signal and of the background.
Double jet $b$-tagging 
reduces the efficiency for detecting the signal but improves 
significantly on the background rejection especially as
regards the largest backgrounds from  $\gamma q \ra g q$
or from 2-resolved $bx$, $cx$ or $\ccbar$ production, where $x$
could be any jet.
 We have generated all final states with at least a
$b$ or a $c$ quark, considering five different possibilities:
\begin{itemize}
\item  perfect $b$-tagging $\epsilon_b=1$, $\epsilon_c=0$,  
$\epsilon_x=0$
 
\item ``optimistic" $b$-tagging efficiency $\epsilon_b=0.6$, 
$\epsilon_c=0.05$, $\epsilon_x=0.002$\\   
Such values for the tagging efficiencies are not available with
present day detectors.  We include them in the hope that they may be 
be realized in a next generation of 
vertex detectors.
\item ``realistic" efficiency $\epsilon_b=0.47$, $\epsilon_c=0.11$,
$\epsilon_x=0.01$    \\
 This is of the order of what can be achieve now at LEP  with  
a microvertex detector. 
\item ``pure" $b$-tagging    $\epsilon_b=0.28$, $\epsilon_c=0.03$,
$\epsilon_x=0.002$    \\
This    $b$-tagging can be obtained  with 
the same  detector as the one used for the ``realistic'' $b$-tagging. The only
difference resides in  the criteria for jet identification 
(requiring 4 rather than 3 tracks with a large
impact parameter) implying a
  lower efficiency for $b$ recognition but  also   a much lower 
contamination level from $c$.  
 \item ``no  $b$-tagging'' $\epsilon_b=1$, $\epsilon_c=1$, $\epsilon_x=0.01$  \\
By this we mean that there is a total confusion between 
$b$ and $c$ only but that the other light flavours and gluons are efficiently
identified.
\end{itemize}

Needless  to say that if
one has no distinction at all between the flavours 
the situation is much worse 
than in the case of ``no $b$-tagging'' (in fact it is hopeless). 
  With a  single-jet tag strategy\cite{eevvh} the ``realistic'' 
$b$-tagging gives a global efficiency of 
$\epsilon_{b\bar{b}}=$0.72, $\epsilon_{c\bar{c}}=$0.21,
 $\epsilon_{b\bar{c}}=$0.53, $\epsilon_{bx}=$0.48, $\epsilon_{cx}=$0.12. 
With double-jet tag the efficiencies change to
$\epsilon_{b\bar{b}}=$0.22, $\epsilon_{c\bar{c}}=$0.01,
 $\epsilon_{b\bar{c}}=$0.05, $\epsilon_{bx}=$0.005, $\epsilon_{cx}=$0.001.
When comparing these numbers the advantage of 
the double-tag over single-tag becomes obvious, especially as regards
the background suppression of all processes containing a $c$ quark.
 Even so, excellent tagging is not 
sufficient, at 500~GeV, to reduce the background from
$gc\ra gc$ below the signal level. To achieve this, cuts
must be implemented.
 The main background that
remains is the 1-resolved $\bbbar$ production as well as the $Z$
 for the lower mass
range. At 350~GeV the dominant background is still the 1-resolved
$\bbbar$ production, although the  2-resolved
$cx$ or $cg$ as well as the direct $\bbbar$
production become comparable around $130$~GeV and $140$~GeV respectively.
Improvement in tagging efficiencies would somewhat
change this picture as regards the relative importance of
processes with $b$ or $c$ quarks.
Note that even the ``optimistic'' efficiencies we are considering are
much less than what is often considered ($\epsilon_{bb}=0.9$). 
 Probably 
by the time (or even before) this machine is built one could achieve 
better efficiencies, but it is not clear how these detectors will 
perform in the \gag\ environment.

\subsection{Cuts}

 From the previous discussion, it should be evident that
cuts on the variables $p_T$ and $p_Z$ would be very useful. In order
to find the best cuts we used  a simple algorithm that chose
 the cuts that would optimize the significance, $\bruit$. 
To calculate the signal ($S$) we have taken events in the invariant mass range
$M_H \pm 2\delta$ where $\delta$ is the resolution. 
Note that, with our definition of the resolution, this means that 
$95\%$ of the signal is contained in a ``box" of 
width $4\delta$. The background ($B$) was
evaluated in the same region. 
For the production of two jets, there are only three independent 
variables:
we used  $M_{jj}$, $p_T$ and $p_Z$.
There is no further need to motivate the choice of the  variable
$p_T$; on the other hand, $p_Z$   was  chosen because it is 
directly related to the boost.  Furthermore, 
 we expect to be able to measure $p_Z$ with roughly the
same precision as $p_T$  for events in  the central 
region (large $p_T$) with little boost. These will 
constitute the bulk of the events once  cuts on 
$p_T$ and $p_Z$  have been applied. 
  Only events where $p_Z$ is large and that
are further away from the central region are expected to 
have a larger uncertainty. Since these events should be far from
the boundary of the region to be cut, they should not cause much problem.  

 \begin{figure*}[htb]
\begin{center}
\caption{\label{scatter}{\em 
Double distribution in the variables $p_Z$-$p_T$ of signal(a) 
and background from
(b) direct $\qqbar$ production
(c) 1-resolved and
(d) 2-resolved $\qqbar$ production. The signal corresponds
to $M_H=120$~GeV, $\delta=5$~GeV and the 
polarizations, $P_c=1,2\lambda_e=0.9$.}}
 \vspace*{-1.cm}
 \mbox{\epsfxsize=15.5cm\epsfysize=15cm\epsffile{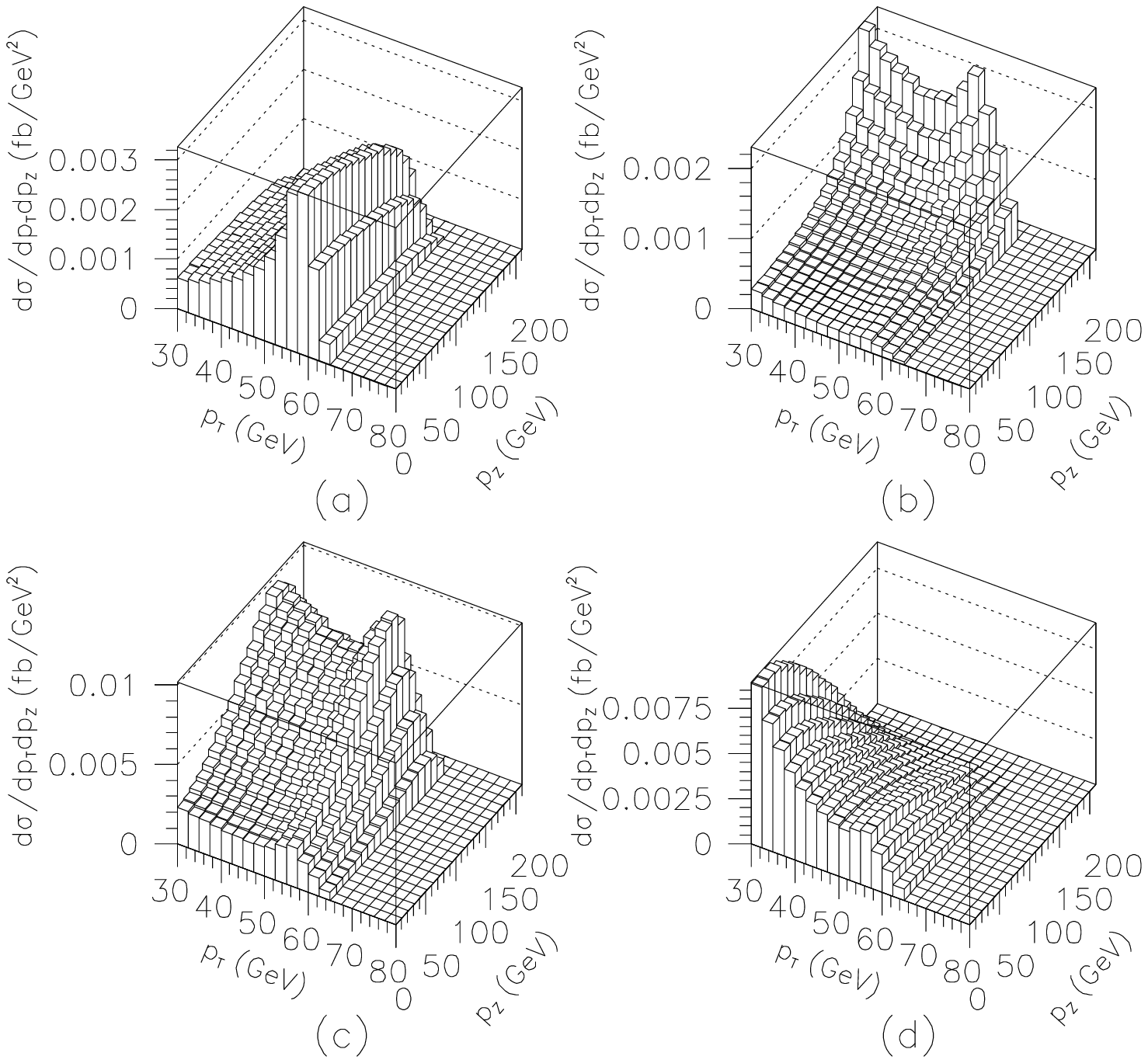}}
 \vspace*{-1.cm}
\end{center}
\end{figure*}

The algorithm works as follows.  First assume that we know the Higgs mass.
We then keep all events such
that $M_{jj}$ is within $2\delta$ of the assumed Higgs mass.
This selects one of the three independent variables so that a scatter
plot of the signal and backgrounds for the two remaining variables 
can be produced.  
The bins were then ordered by increasing value of $S/B$. 
Starting with the bin having the best $S/B$, 
all the bins were then added in decreasing order.   
 At each iteration $\bruit$ was calculated. The summation was continued  until the 
significance given by $\bruit$ started to decrease.
This defined the region to be kept. The procedure
  was then automatized to calculate optimal cuts 
 for each particular case corresponding to 
different Higgs masses, as well as to the parameters of the experiment
considered:
single or double-tag, tagging efficiencies, resolution of the
apparatus and polarization of the initial beams.
With the optimal cuts, we  then estimated the $\bruit$ for each case.
Considering the fact that the $M_{\bbbar}$ interval where we evaluate both the
signal and background is set by the resolution, 
  a better resolution  reduces  the
background significantly (roughly by a factor of 2 from 10 to 5~GeV resolution).
In Figure~\ref{scatter}, we show the double distribution in the 
$(p_T,p_Z)$ variables at $\sqrt{s_{ee}}=500$~GeV  for 
direct, 1-resolved and 2-resolved  backgrounds, as well as for the signal. 
We have taken $M_{b\bar b}=120\pm 2\delta$~GeV, $\delta=5$~GeV, 
 a ``realistic'' $b$-tagging efficiency and a good polarization 
$P_c=1,2\lambda_e=0.9$. 
The characteristics of the distributions are similar for 
the lower energy collider.
The signal shows up as a peak in the $p_T$ distribution
while the
direct background concentrates at high $p_Z$ just as
the 1-resolved one. The latter also predicts a large number
of events at small $p_T$ and constitutes the dominant background.
The 2-resolved events gather in the small $p_T$ region and form
the second largest background. However, this background
 can be effectively eliminated
with a cut on $p_T$. This cut  should also
reduce  the 1-resolved background while cutting the high values of
$p_Z$  is essential to  reject both  the direct and the 1-resolved.
This is precisely what we find with the algorithm to optimize the cuts.

\begin{figure*}[htb]
\begin{center}
\caption{\label{cuts}{\em 
Optimal cuts in the
$p_T-p_Z$ plane for 
(a) $M_H=100$~GeV, $2\lambda_e=0.9$
(b) $M_H=120$~GeV, $2\lambda_e=0.9$
(c) $M_H=140$~GeV, $2\lambda_e=0.9$
(d) $M_H=120$~GeV, $\lambda_e=0$.
Full line is for $500$~GeV,
dashed one for $350$~GeV\@.}}
 \vspace*{-1.cm}
 \mbox{\epsfxsize=15.5cm\epsfysize=15cm\epsffile{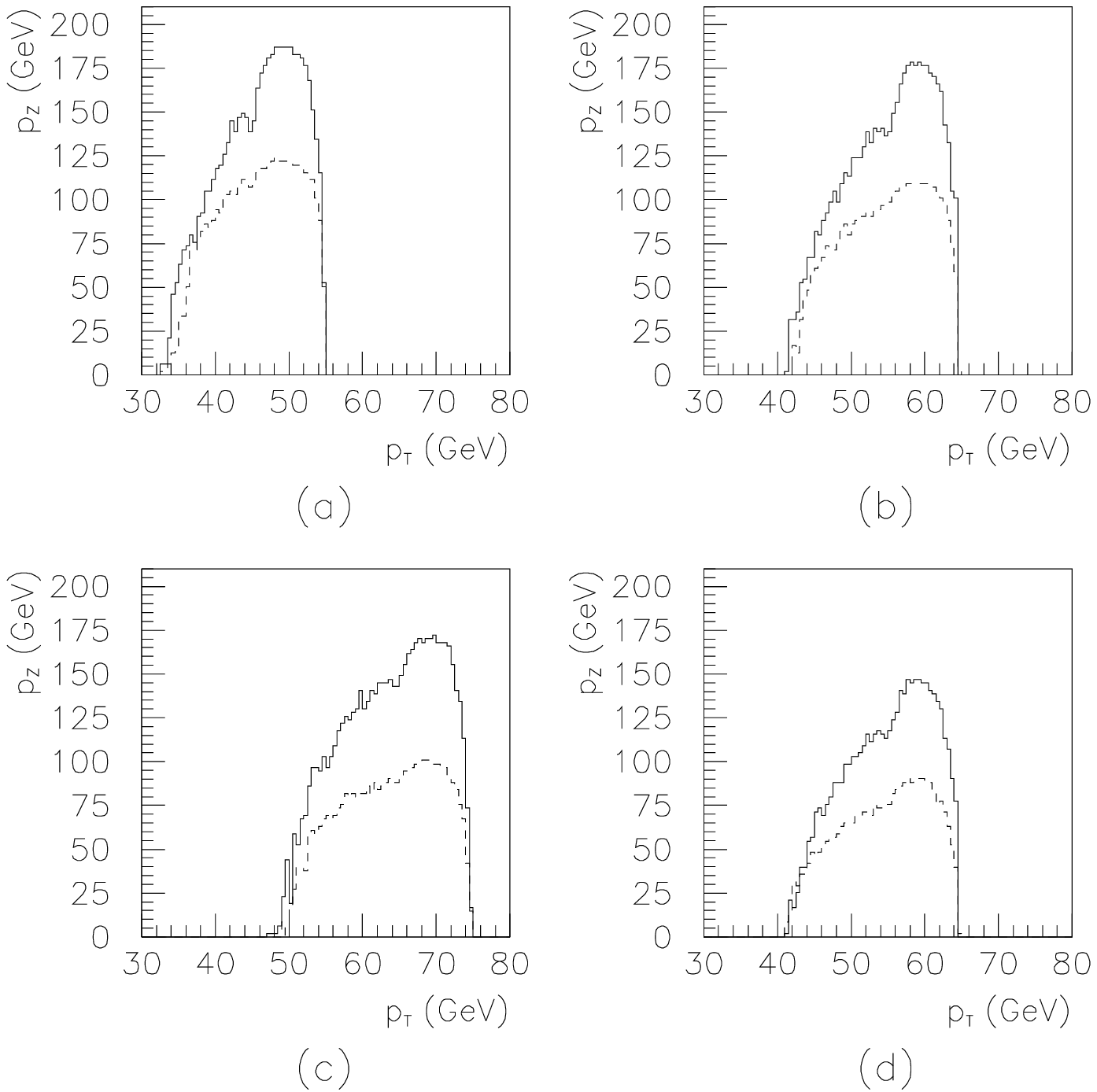}}
 \vspace*{-1.cm}
\end{center}
\end{figure*}

The results of the optimization procedure are shown in 
Fig.~\ref{cuts} where the region in the $p_T$--$p_Z$ plane containing the
events to be kept are presented for both the
500~GeV and 350~GeV colliders and for different masses. ``Realistic'' tagging,
optimal laser polarization and 5~GeV mass resolution are assumed.
 Although the optimal region varies  as a function
of the set-up used and of  $M_H$ (or $M_{\bbbar}$), 
 the optimal cut on $p_T$ corresponds roughly to 
\beqn
\label{pt}
\max(30~\mbox{\rm GeV}, 0.375 M_{\bbbar}) < p_T <
\frac{1}{2}(M_{\bbbar}+2\delta).
\eeqn
The $p_Z$ cut depends on both $p_T$ and on the region 
in $M_{b\bar{b}}$  considered, the upper limit decreasing with 
$M_{b\bar{b}}$.
 It might seem cumbersome to optimize the region where cuts should
be applied. 
We point out that, in fact, a ``rectangular cut'' in the space of the
two variables $p_T$, as in eq.~\ref{pt},  and 
$p_Z$ as given in Table~\ref{pzcut}, is almost
as effective. By rectangular we mean fixed and uncorrelated values 
for $p_T$ and $p_Z$, imposed only from a knowledge (or assumption) on $M_H$. 
Note also that, in this case and for all set-ups, 
the $p_Z$ cut scales as $M_H$ and thus can be 
inferred from Table~\ref{pzcut}. 
For example with ``realistic'' double jet tagging at $500$~GeV,
good resolution and the polarized spectrum, the $\bruit$ for $M_H=$120 (140)~GeV
goes from 4.0 (4.4) to 3.8 (4.2) in going from the optimal cuts to
the rectangular  cuts. In all cases the variations
in $\bruit$ were at most 5\%.

 \begin{table*}[htb]
\caption{\label{pzcut}
 {\em  Cuts on $p_Z$ (in GeV) for different Higgs mass and settings.
Except for the unpolarized case, the laser polarization
is assumed maximal, $P_c=1$.}}
\begin{center} 
\vspace*{0.3cm}
\begin{tabular}{|ccc|cc|cc|} \hline 
&&&\multicolumn{4}{c|}{$p_Z$} \\
\cline{4-7}
Polarization & $\delta$ & tagging  & \multicolumn{2}{c|}{$\sqrt{s_{ee}}=500$~GeV}&
\multicolumn{2}{c|}{$\sqrt{s_{ee}}=350$~GeV} \\
$2\lambda_e$ & (GeV) &           &$M_H$=120 &$M_H$=140 &$M_H$=100 &$M_H$=140 \\ \hline
    0.9       &   5   & realistic & 155 & 150 & 115 & 90  \\
    0.5       &   5   & realistic & 145 & 135 & 100 & 75  \\
     0       &   5   & realistic & 140 & 130 & 90  & 70  \\
   unpol     &   5   & realistic & 190 & 190 & 125 & 110 \\
   0.9        &  10   & realistic & 160 & 155 & 110 & 90  \\
   0.9        &   5   &  no tag   & 155 & 150 & 105 & 85  \\ \hline
\end{tabular}
\end{center}
\end{table*}

In the final analysis we implemented 
the cuts obtained from the optimization algorithm.
These  gave marginally better results. 

If the Higgs mass had not been determined previously, 
the cuts would be implemented in a similar way. 
The procedure would only be more complicated since 
one would be forced to do a scan over the whole range
of values. A search for the Higgs signal would 
then be performed, as described, for 
each particular assumed value of the  mass.

\section{Results}

From the previous discussion on the background and its rejection,
we would expect that the optimal conditions for
observing the IMH would be with 
a ``low-energy" collider using polarized spectra and a double-tag 
strategy. These expectations are confirmed by a detailed analysis
of the discovery potential of both  500~GeV and 350~GeV colliders.
A comparison of significance levels constitutes the 
basis for discussing the merits of the various  set-ups.
Most of our discussion on the discovery potential is 
based on a canonical value of 10~fb$^{-1}$ for the integrated luminosity.

\subsection{The case of a 500~GeV \boldmath{\epm}}
\begin{figure*}[p]
\begin{center}
\vspace*{-1.cm}
\caption{\label{mbb500}{\em The fate of the Higgs resonance at $500$~GeV 
as the polarization of the beams is varied. 
The combined effect of the resolution ($\delta$) 
and the $b$-tagging efficiencies is shown. 
The two peaks correspond to $M_H=120$ and $140$~GeV\@.
The dashed line shows the
resolved contribution. }}
\vspace*{-1.cm}
\mbox{\epsfxsize=17.cm\epsfysize=23cm\epsffile{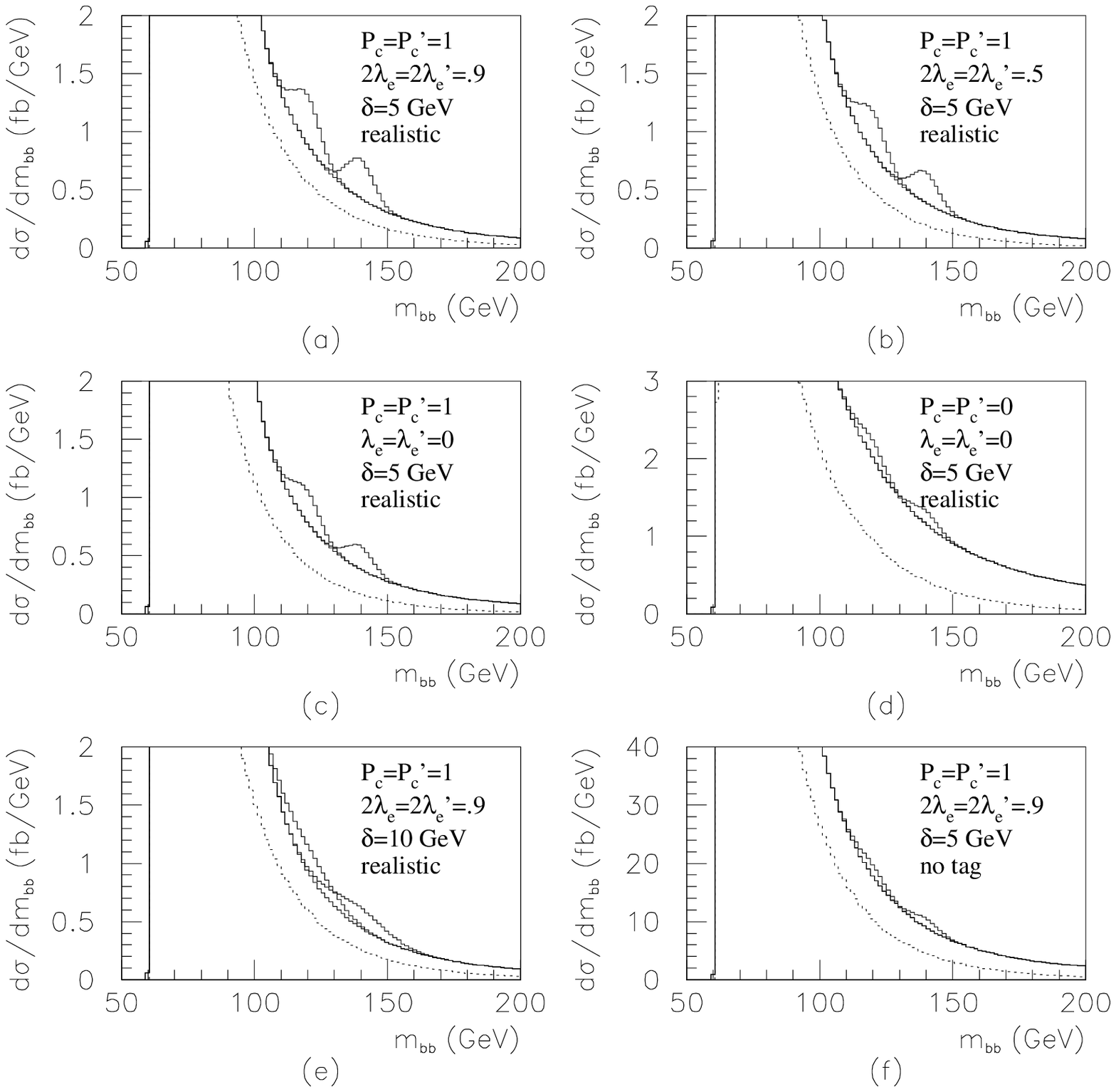}}
\vspace*{-6.cm}
\end{center}
\end{figure*}
\subsubsection{Overview}
Before entering in the detailed influence of each of the parameters of the 
collider (polarization) and of the detector (tagging efficiencies and 
mass resolution) we begin by illustrating the main effects brought about 
by a change in these parameters. This should help when we proceed 
to the systematic analysis of the observability of the Higgs signal 
and the choice of these parameters. Here the double-tag strategy is assumed.

The
effect of a loss in  polarization on both the signal
and background is shown in Figure~\ref{mbb500}a-d. By taking 
the ``realistic" tagging efficiencies together with ``our best" resolution 
($\delta=5$~GeV), the signals 
corresponding to a  Higgs of $120$~GeV or $140$~GeV,
slowly disappear as the degree of polarization degrades. 
The cuts imposed are the
``rectangular'' ones of Table~\ref{pzcut} and eq.~\ref{pt}.
The message from these figure is that,  in order to 
see a peak over the background, some degree of polarization is 
vital. Nonetheless, although one should strive for the
best degree of polarization possible, a modest 
polarization ($2\lambda_e=0.5$) is enough
to guarantee a signal at $\bruit>3$ with the ``realistic'' $b$-tagging 
and with a
 resolution $\delta=5$~GeV\@. The important point to stress for the case of the 
$500$~GeV collider is that, having a larger $J_Z=0$ helps in 
enhancing the signal rather than to further reduce the background that is 
dominated by the resolved contribution (see ~Figure~\ref{mbb500}),
 especially after the cuts have been applied. Indeed, these cuts 
have the added advantage of drastically reducing the $J_Z=2$ component 
of the spectrum leading to a very good $J_Z=0/J_Z=2$ ratio. 
The reason this is so and that some loss in the electron polarization is 
not so critical, 
provided the photon is fully polarized, is to be found in section~2 
where we discussed the characteristics of the 
polarized spectra at some length. 
As explained there, the effect of the $p_Z$ cuts is to 
filter photons with sensibly the same energy and therefore, provided 
the original laser photons have like-sign helicities, 
the colliding photons inherit 
almost the same degree of polarization, which leads to a spectrum 
that is $J_Z=0$ dominated (see
Fig.~\ref{spectre34}b). 
Of course, this also improves the 
rejection of the direct processes. 
Still,  the magnitude of the $J_Z=0$ luminosity is higher with an 
increased electron polarization,
hence the advantage of polarizing the electron as much as possible.

The change in the resolution from $5$~GeV to $10$~GeV 
has  a  dramatic effect: the nice peak structure has almost
disappeared in Fig.~\ref{mbb500}e and the significance, for $M_H=$120 (140)~GeV, 
goes from $\bruit=3.8 (4.2)$ with the better resolution 
($\delta=5$~GeV) to $\bruit=2.5 (3.0)$.
One should, however, be careful with a qualitative pictorial analysis of the 
signal based on Figure~\ref{mbb500}. These figure could be slightly misleading
as they only show the signal relative to the background.
Two set-ups that give the same $S/B$ could have different significance
level ($\bruit$), a larger absolute number of events for the signal leading
to a better significance. When comparing two seemingly
identical figures, note should be made of the absolute scale.
For example, in Fig.~\ref{mbb500}f one sees very small peaks
for the Higgs in the case of no-tagging; however the $\bruit$
are practically the same as  the ``realistic'' tagging of
Fig.~\ref{mbb500}a, all other parameters chosen at the same value.
In the following, we use the 
  significance levels to compare  various tagging strategies and
efficiencies.

\subsubsection{Tagging strategies and detector performances}
In Table~\ref{sig500}, the significance  for  maximal 
polarization of the laser and $2\lambda_e=0.9$ are given using optimal
cuts. The first remark is that 
 double tagging is much better than single tagging. This can be traced 
back to the fact that with double tagging,
one can practically eliminate the 1-resolved background 
$\gamma q \ra g q$ while   improving on the rejection of  
$\gamma g \ra q \bar{q}$. The latter is, however, 
 done at the expense of a reduction in the signal. Nevertheless,
 the overall improvement is significant.
With the double-tag strategy, which we have just established
as being the most favourable one, an 
  unexpected result  stands out:  one can make do with 
``no tagging''  (meaning total confusion between $c$ and $b$ only).
Indeed,   the same significance levels can be reached whether one uses 
  the so-called ``realistic'' $b$-tagging or ``no-tagging''. At first, this result 
might seem  surprising, but it can be understood simply by estimating 
the effect of losing the tagging on the signal 
(increasing it by $\approx 4$) and on the 
1-resolved $\qqbar$ production (roughly increasing the background by 16).  
This is not to say that tagging is not useful; one has to weigh the benefit 
of ``selection" (filtering the signal through tagging)
against that of statistics (keeping enough signal events).
With the rejection power 
in the ``realistic'' tagging scenario one loses so much in 
statistics that the significance is at almost the same level as in the
case of ``no $b$-tagging''. In our case tagging will only be beneficial if an improved 
detector is available. 
We would not have reached this conclusion had we only
considered the single-tag strategy. In that case, any type of tagging
 is useful especially as it   improves
on the rejection of  1-resolved production of $c$ quarks
which contribute to the main background. For the same reason,
improving on the purity of the tagging (better rejection
of $c$ quarks) is of some help. On the other hand,  
 with a    double-tag strategy, there are no advantages
in  striving for the ``purer" tagging since the improved  background reduction 
  is offset by a sharp drop in the signal. 

\begin{table*}[htb]
\caption{\label{sig500}
{\em  Significance levels for IMH at
500~GeV with polarized spectrum, $P_c=1,\lambda_e=0.9$
 and various resolutions and tagging.
Optimal cuts are used. $\int\! \cal{L} 10$~fb$^{-1}$.}} 
\vglue.1in
\begin{center}
\begin{tabular}{|ccc|ccccc|ccccc|}\hline 
&&&&
\multicolumn{3}{c}{double-jet tag}&&&
\multicolumn{3}{c}{single-jet tag}&\\

$M_H$&$\delta$&$\epsilon_b$&1&.6&.47&.28&1&1&.6&.47&.28&1\\
(GeV)&(GeV)&$\epsilon_c$&0&.05&.11&.03&1&0&.05&.11&.03&1\\
&&$\epsilon_x$&0&.002&.01&.002&0.01&0&.002&.01&.002&.01\\\hline

90&5&&3.1&1.8&1.2&0.8&1.4&1.9&1.5&1.2&1.2&0.8\\
90&10&&2.4&1.4&0.9&0.6&1.0&1.4&1.1&0.9&0.9&0.5\\\hline

100&5&&4.8&2.8&1.9&1.3&2.0&2.8&2.3&1.8&1.8&1.1\\ 
100&10&&3.5&2.0&1.4&1.0&1.5&2.0&1.7&1.3&1.3&0.8\\\hline 
 
120&5&&10.6&6.1&4.0&2.8&4.0&5.5&4.5&3.3&3.5&2.1\\ 
120&10&&7.1&4.1&2.7&1.9&2.8&3.9&3.1&2.3&2.4&1.4\\\hline
 
140&5&&12.2&7.0&4.5&3.2&4.4&6.3&5.0&3.7&3.9&2.2\\ 
140&10&&8.6&4.9&3.2&2.3&3.1&4.4&3.5&2.6&2.7&1.6\\\hline
\end{tabular}
\end{center}
\end{table*}

\begin{figure*}[htb]
\begin{center}
\vspace*{-1.cm}
\caption{\label{opt500}{\em The Higgs resonance and its background 
at $500$~GeV 
assuming ``realistic''   $b$-tagging and  a resolution $\delta=5$~GeV for
(a) $M_H=120$~GeV   (b)  $M_H=140$~GeV\@.
Full lines are with the optimal (automatized) set of cuts 
while dotted lines are for the fixed $p_T$-$p_Z$ cuts (see text).}}
\vspace*{-1.5cm}
\mbox{\epsfxsize=17.cm\epsfysize=11cm\epsffile{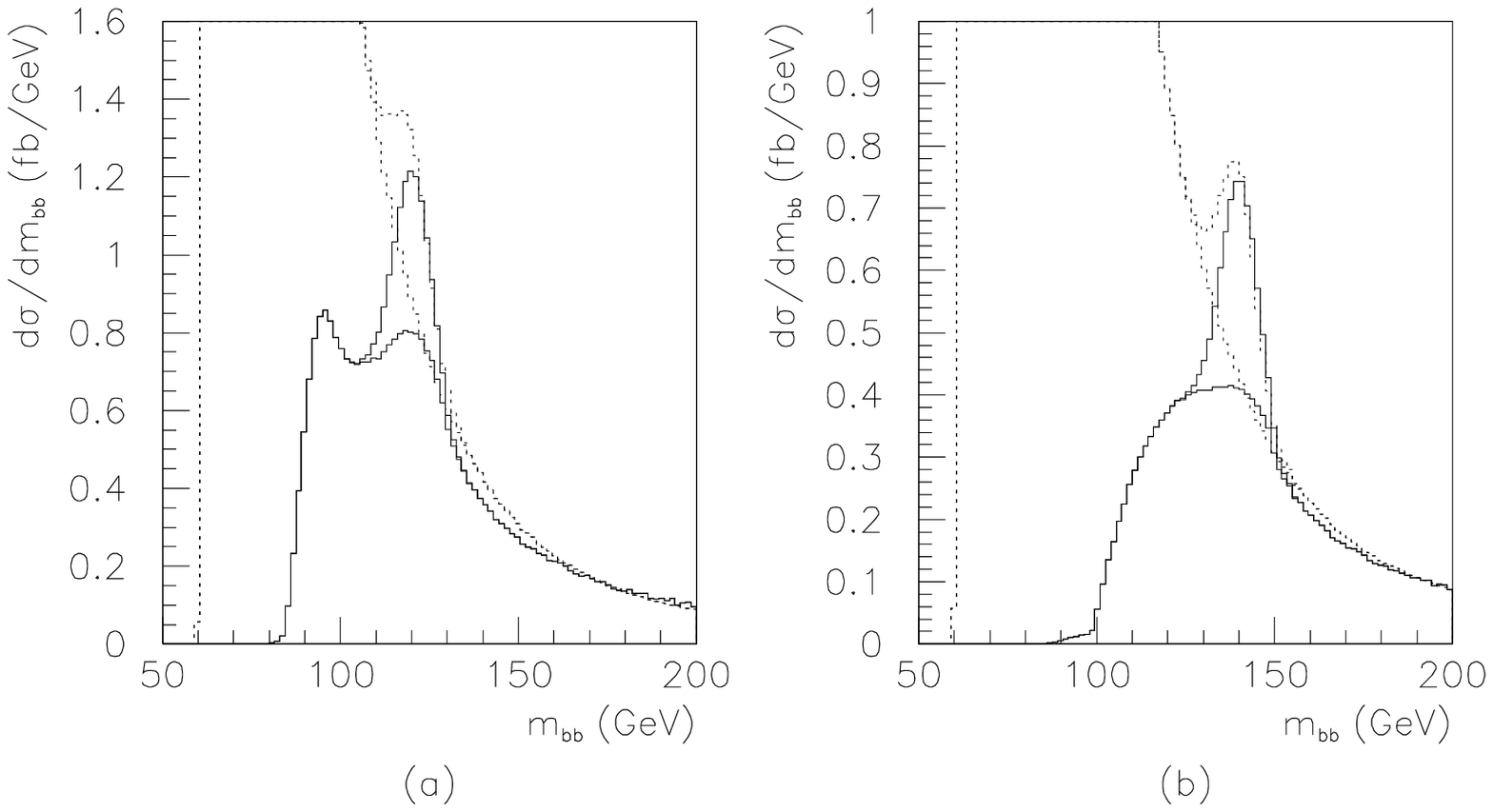}}
\vspace*{-1.cm}
\end{center}
\end{figure*}

In the following, only results corresponding to the best tagging 
strategy, the double-tag,
 will be quoted. An integrated luminosity of only
10~fb$^{-1}$ is assumed. With a good resolution (5~GeV) and a ``realistic'' 
$b$-tagging  together with $90\%$  longitudinal polarization 
for the electrons, we obtain a good signal with a significance 
$\sigma=S/\sqrt{B}=4.0$ ($\sigma=4.5$) for $M_H=120$~GeV ($M_H=140$~GeV).
{\footnote{These numbers differ from the ones we have already presented
\cite{Paris} 
due mainly to the addition of  2-resolved processes and $\gamma q \ra gq$
and to the different tagging strategy used.}} 
The signal and background for these two masses are shown in 
Fig.~\ref{opt500} after optimal cuts have been applied. The comparison 
with using the ``rectangular", {\it i.e.}, uncorrelated $p_T$-$p_Z$ cuts,
is also shown. 
The latter cuts reproduce the more familiar 
continuum background. Note that the optimal cuts for $M_H=120$~GeV also reveal 
the $Z$ resonance. 

 With the  luminosity considered and the ``realistic'' (present-day performance) 
$b$-tagging
efficiencies,
regardless of the resolution or polarization chosen, it is impossible to
detect a 90 or 100~GeV Higgs at a 500~GeV collider.
For this to be possible, without luminosity increase, 
requires a near perfect tagging efficiency.
This is due in part to the presence of the $Z$ and in part to 
the higher level of resolved background.
Since the significance will increase like $\sqrt{{\cal L}}$,
one can estimate the necessary luminosity
 to see a signal at the $3\sigma$   level. 
We find that a luminosity of 45 (60)~fb$^{-1}$ will be
needed to see a 100 (90)~GeV Higgs with a resolution of 5~GeV while
30 (50)~fb$^{-1} $ will suffice if the ``optimistic" tagging is available.
A signal at   the $5\sigma$ level calls for a very high luminosity
 75 (110)~fb$^{-1} $ for $M_H=100 (90)$~GeV\@. 
This is to be compared with the 11 (15)~fb$^{-1}$  necessary to obtain
a 5$\sigma$ effect  for $M_H=120 (140)$~GeV\@.

\vspace*{1.5cm}

\subsubsection{The issue of resolution}
Unless one has an improved microvertex detector corresponding to 
the ``optimistic" scenario operating with a double $b$-tag 
strategy, with a change in the resolution from $5$~GeV to $10$~GeV 
one can barely see the 140~GeV Higgs,
let alone the $120$~GeV, even when the electron polarizations are at $90\%$. 
In the case where only the `realistic'' tagging is possible, the
significances drop from 4. to 2.7 for $M_H=$120~GeV and from 4.5 to 3.2 for
$M_H=$140~GeV\@.
The loss in resolution, however, can be almost exactly
compensated for by an improved $b$-tagging efficiency.  Compare
in Table~\ref{sig500} the figures for $\delta=5$~GeV, $\epsilon_b$=0.47
(``realistic'') with
the ones for $\delta=10$~GeV, $\epsilon_b$=0.6 (``optimistic''). 

\begin{table*}[htb]
\caption{\label{sig500pol}
{\em   Significance levels for IMH at
500~GeV\@. Effect of polarization of the electron.
Except for the unpolarized case the laser is assumed to 
be perfectly polarized. The expected number of events are shown 
in parenthesis for the best polarization $\delta=5$~GeV\@. The tagging efficiencies
used are the same as in the previous table. Optimal cuts are
used.}}
\vglue.1in
\begin{center}
\begin{tabular}{|c|c|c|c|c|c|c|}\hline 
  Mass&$\delta$&$b$-tag&$2\lambda_e=0.9$&$2\lambda_e=0.5$&
$2\lambda_e=0$&unpol\\
(GeV)&(GeV)&&&&&\\\hline
120& &0.6&6.1(81)&5.2&4.3&2.1\\
120&5&0.47&4.0(50)&3.4&2.7&1.3\\
120& &no&4.0(215)&3.4&2.7&1.1\\\hline

120& &0.6&4.1(78)&3.5&2.9&1.4\\
120&10&0.47&2.7(47)&2.3&1.9&0.9\\
120& &no&2.8(206)&2.4&1.9&0.8\\\hline 

140& &0.6&7.0(66)&5.8&4.4&2.1\\
140&5&0.47&4.5(40)&3.7&2.8&1.4\\
140& &no&4.4(174)&3.5&2.6&1.1\\\hline

140& &0.6&4.9(63)&4.0&3.1&1.5\\
140&10&0.47&3.2(38)&2.6&2.0&1.0\\
140& &no&3.1(166)&2.5&1.9&0.8\\\hline
\end{tabular}
\end{center}
\end{table*}
\subsubsection{Polarization}
We have already qualitatively discussed the importance of
polarization of the beams in Fig.~\ref{mbb500}a-d.
A more detailed  analysis leads to the
 significance levels of Table~\ref{sig500pol}
 in the double-tag strategy for the two ``visible" masses,
120 and 140~GeV\@.
The previous conclusion stands:
some degree of polarization is vital. We cannot even afford to 
polarize only the lasers; if this is the case a good significance
is achieved only with the ``optimistic'' tagging 4.3 (4.4)
for 120 (140)~GeV and the best resolution $\delta=5$~GeV\@.
However, a modest electron polarization (50\%) would be sufficient
to have a signal at $\bruit>3$ without $b$-tagging and with a
 resolution $\delta=5$~GeV\@.
In table~\ref{sig500pol}, we also give  the number of events expected 
for  the best polarization. Notice the much larger number of events
in the absence of tagging for the heavy quarks compared to the
``realistic'' tagging. These can be compared since the cuts used in the
two cases are similar.

To conclude, we stress the primary 
importance of having  a good resolution. 
One could even make do without any tagging with the exception of identifying 
the light quarks. Although   $b$-tagging is not a 
critical issue, as long as a double-jet strategy is used,
  a $b$-tagging better than the one that can be achieved nowadays (at LEP)
would be profitable; it could even compensate for a loss in resolution.
  The polarization settings seem to be 
achievable without much problem since an excellent degree of $e^-$ 
longitudinal 
polarization is not absolutely essential, though helpful.
We stress that it is vital to polarize the lasers.

Although we have not discussed the case of a 110~GeV Higgs,
it is easy to evaluate  its
significance from fitting the results obtained
for the four Higgs masses that we studied in detail.
For example,
with a mass resolution of 5~GeV and 90\% electron
polarization, a conservative estimate gives a significance
larger than three.  As a much lower Higgs mass suffers significantly 
from the $Z$ background for a 5~GeV mass resolution,
$M_H\approx110$~GeV can be considered the lowest Higgs mass
accessible by a 500~GeV collider.

\subsection{The case of a 350~GeV \boldmath{\epm}}

\begin{figure*}[p]
\begin{center}
\vspace*{-1.cm}
\caption{\label{mbb350}{\em As in 
Fig.~\protect\ref{mbb500} but for the case of the 
350~{\rm GeV} collider.}}
\vspace*{-1.cm}
\mbox{\epsfxsize=17.cm\epsfysize=23cm\epsffile{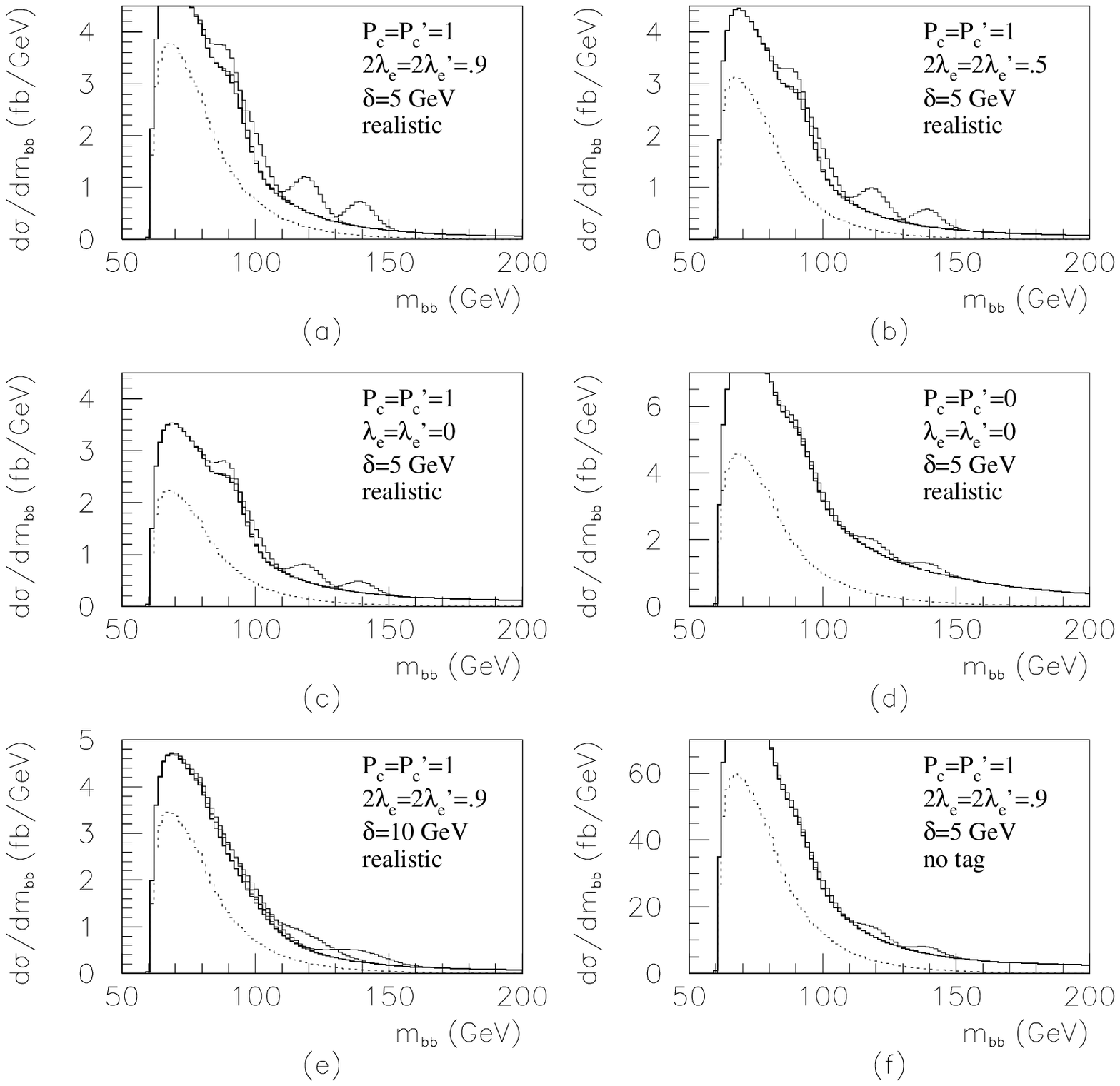}}
\vspace*{-1.cm}
\end{center}
\end{figure*}

Here the situation is far better: see Fig.~\ref{mbb350} where
the four peaks corresponding to $M_H=90,100,120$ or $140$~GeV
dominate over the background.
This is mainly because the resolved contributions have dropped.
In these figures, 
the ``rectangular" cuts of eq.~\ref{pt} and Table~\ref{pzcut} are
used.
As was the case for the higher energy machine, a loss in 
polarization reduces the signal, see Fig.~\ref{mbb350}a-d, and
polarization of at least the laser beams is absolutely 
essential.
Achieving a good resolution is also very useful: with $\delta=10$~GeV,
$2\lambda_e=0.9$ and ``realistic'' tagging,
the signal disappears for $M_H=100$~GeV, with $\bruit=2.6$.
Although the peaks are not as prominent in the
absence of tagging for the heavy quarks (Fig.~\ref{mbb350}f), the
larger absolute number of signal events compensate partially for that
and the significances are only slightly 
smaller than the ones for the ``realistic'' tagging. 
For example, for $M_H=120$~GeV 
we get $\bruit=7.3$ for no-tag instead of 7.8
for ``realistic'' tagging. 

\begin{table*}[htb]
\caption{\label{sig350}
{\em Significance levels for IMH at
350~GeV with polarized spectrum and various resolution and tagging.}}
  \vspace*{0.3cm}
\begin{center}
 \begin{tabular}{|ccc|ccccc|ccccc|}\hline 
&&&&
\multicolumn{3}{c}{double-jet tag}&&&
\multicolumn{3}{c}{single-jet tag}&\\
 
$M_H$&$\delta$&$\epsilon_b$&1&0.6&0.47&0.28&1&1&0.6&0.47&0.28&1\\
(GeV)&(GeV)&$\epsilon_c$&0&0.05&0.11&0.03&1&0&0.05&0.11&0.03&1\\
&&$\epsilon_x$&0&0.002&0.01&0.002&0.01&0&0.002&0.01&0.002&0.01\\\hline
 
90&5&&6.6&3.8&2.6&1.8&2.8&3.9&3.2&2.4&2.5&1.5\\
90&10&&5.1&2.9&2.0&1.4&2.0&2.9&2.3&1.7&1.8&1.1\\
 
100&5&&10.2&5.9&4.0&2.7&4.2&5.9&4.8&3.5&3.7&2.2\\
100&10&&7.5&4.3&2.9&2.0&3.0&4.3&3.4&2.5&2.7&1.5\\
 
120&5&&22.7&12.8&8.1&5.9&7.6&11.5&8.9&6.4&6.9&3.8\\
120&10&&15.0&8.5&5.5&3.9&5.3&8.0&6.2&4.4&4.8&2.7\\
 
140&5&&24.7&13.9&8.5&6.4&7.5&12.5&9.3&6.6&7.3&3.9\\ 
140&10&&17.3&9.7&6.0&4.4&5.3&8.8&6.5&4.6&5.1&2.7\\\hline
\end{tabular}
\end{center}
\end{table*}

For the detailed comparison of the significances for 
various polarization set-ups, tagging strategies and efficiencies, 
we again prefer to use the optimal cuts
 obtained with the procedure described in section 4.
The results with the best polarization ($2\lambda_e P_c=0.9$) are given in 
Table~\ref{sig350}. 
As for the higher energy machine and for similar arguments,
double tagging is always much better.    
 With $90\%$ longitudinal electron 
polarization and 5~GeV resolution we obtain good significance levels 
  for $M_H \ge 100$~GeV\@. With the ``realistic" efficiencies
we have $\sigma=4.,8.1$ and $8.5$ for $M_H=100,120$ and $140$~GeV
respectively. For the last two values, this leads to
an unequivocal signal and possibly enough events to do precise measurements
of the $H\gamma\gamma$ coupling (see below). 

 Taking a larger resolution (10~GeV) the signal still has a   
statistical significance $\sigma\ge 3$ for $M_H\ge 100$~GeV,
 even without $b$-tagging,   if a very good degree 
of longitudinal polarization for the 
$e^-$ ($2\la_e=2\la_e'=90\%$) can be achieved.
Even at this energy it will be hard to obtain a meaningful
signal for a IMH of $90$~GeV\@. For this, optimal settings are needed.
Indeed, with the ``realistic" tagging, optimal polarization 
$2\lambda_e P'_c=2\lambda_e P_c'=0.9 $ and
 resolution of $\delta=5$~GeV, we get $\sigma=2.6$. The 
minimum luminosity to have $\sigma\ge3 (5)$ is 13 (35)~fb$^{-1}$.
With the nominal luminosity of 10~fb$^{-1}$, only  ``optimistic'' tagging 
gives a good significance, $\bruit=3.8$.

\begin{table*}[tp]
\caption{\label{sig350pol}
{\em  Significance levels for IMH at
350~GeV: Effect of polarization of the electron.
Except for the unpolarized case the laser is assumed to 
be perfectly polarized. Double jet-tag strategy and optimal
cuts are used. The expected number of events is shown 
in parenthesis for the best polarization. $\int \cal L=10$~fb$^{-1}$ is 
assumed.}}
\vglue.1in
\begin{center}
\begin{tabular}{|cc|c|c|c|c|c|}\hline 
  Mass&$\delta$&$b$-tag&$2\lambda_e=0.9$&$2\lambda_e=0.5$&
$\lambda_e=0$&unpol\\\hline

90& &0.6&3.8(102)&3.2&2.5&1.4\\
90&5&0.47&2.6(62)&2.2&1.7&0.9\\
90& &no&2.8(254)&2.3&1.8&0.8\\\hline

90& &0.6&2.9(92)&2.5&1.9&1.0\\
90&10&0.47&1.9(55)&1.6&1.3&0.7\\
90& &no&2.0(229)&1.7&1.3&0.6\\\hline

100& &0.6&5.9(121)&4.8&3.7&2.0\\
100&5&0.47&4.0(71)&3.2&2.4&1.3\\
100& &no&4.2(291)&3.3&2.5&1.1\\\hline

100& &0.6&4.3(110)&3.5&2.7&1.4\\
100&10&0.47&2.9(64)&2.3&1.8&0.9\\
100& &no&3.0(271)&2.4&1.7&0.8\\\hline
    
120& &0.6&12.8(129)&9.8&6.8&3.4\\
120&5&0.47&8.1(79)&6.1&4.2&2.1\\
120& &no&7.6(335)&5.5&3.7&1.7\\\hline 

120& &0.6&8.5(122)&6.5&4.5&2.4\\
120&10&0.47&5.5(74)&4.1&2.9&1.5\\
120& &no&5.3(322)&3.8&2.6&1.2\\\hline

140& &0.6&13.9(97)&9.6&6.1&3.3\\
140&5&0.47&8.5(58)&5.9&3.7&2.0\\
140& &no&7.5(248)&4.9&3.0&1.5\\\hline

140& &0.6&9.7(94)&6.7&4.2&2.3\\
140&10&0.47&6.0(57)&4.1&2.6&1.4\\
140& &no&5.3(246)&3.4&2.1&1.1\\\hline
\end{tabular}
\end{center}
\end{table*}

A loss in polarization will  again lead to a degradation of the
significance levels (see Table~\ref{sig350pol})
and the effect will be much more dramatic than for the
higher energy collider. 
Nevertheless, the significance levels
  are still good  for $M_H=120$--$140$~GeV,  with no electron 
 polarization, provided one has a good resolution. 
   In this non-optimal case and
with the ``realistic'' tagging,  we obtain $\sigma=4.2$ (3.7) for $M_H=120$ (140)~GeV 
(with $\int \cal{L}=10$~$^{-1}$ only).
With the ``realistic'' tagging, the 10~GeV resolution
and no electron polarization, the significance for the 120~GeV Higgs 
is $\sigma=2.9$ only. 
Note that in this case, due to the increased importance
of the direct background at higher masses, the $140$~GeV IMH 
is undetectable.
Unfortunately, even 
at $350$~GeV, the IMH is lost if ``no $b$-tagging'' is provided and if the resolution 
is large without much electron polarization. But this is the most pessimistic 
scenario.

\begin{figure*}[htb]
\begin{center}
\vspace*{-1.cm}
\caption{\label{opt350}
{\em The Higgs resonance and its background 
at 350~{\rm GeV}
assuming ``realistic''   $b$-tagging and  a resolution $\delta=5$~{\rm GeV} for
(a) $M_H=100$~{\rm GeV}   (b)  $M_H=140$~{\rm GeV}.
Full lines are with the optimal (automatized) set of cuts 
while dotted lines are for the fixed $p_T$-$p_Z$ cuts (see text).}}
\vspace*{-1.5cm}
\mbox{\epsfxsize=17.cm\epsfysize=11cm\epsffile{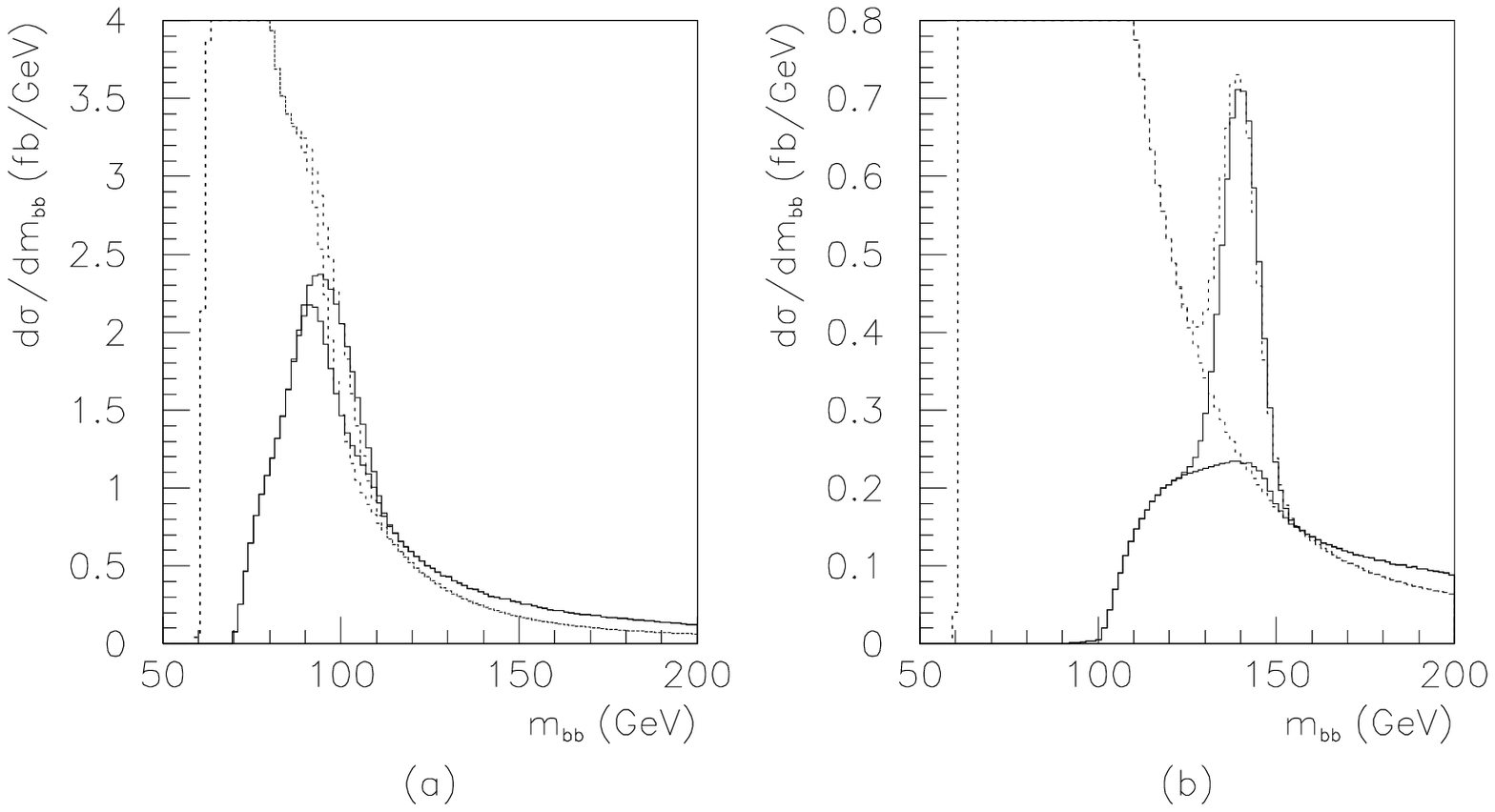}}
\vspace*{-1.cm}
\end{center}
\end{figure*}



\section{Conclusion}

  The purpose of our study has been to critically inquire whether 
the IMH resonance can clearly stand out  in a high-energy $\gam$ collider, 
obtained from a 350~GeV or 500~GeV \epm\ linear machine, if it were 
operated in a broad spectrum. In such a scheme, with $\gam$ cms energies 
extending to a maximum of $\sim80\%$ of the \epm\ energy, one would not 
have to sacrifice carrying out an extensive 
physics programme in a novel environment. 
Our results indicate that the requirements on the  
detector performances and the polarization of the beams are quite 
different for the 350~GeV and 500~GeV machines. One common feature, though, 
is that circular laser polarization, which should be readily 
available, is essential at both energies.  It is important to realize 
that at 350~GeV, $e^-$
polarization is also a top priority while we could, in this case, 
make do with a
not so good resolution, $\delta=10$~GeV, on the invariant $b \bar b$ 
mass. 
On the other hand, at 500~GeV, and with the canonical luminosity 
of $\int {\cal L}=10$~fb$^{-1}$, we could not, for any value of 
the mass of the Higgs, afford
to have such values of the resolution unless $b$-tagging efficiencies
are at least at the level of what we have called ``optimistic''
efficiencies, that is, much better than what is achieved with present 
microvertex detectors working in a clean environment like LEP. 
However, at 500~GeV one can survive with a 
non-optimal longitudinal polarization
of the electron. The reason is that the main background at the higher 
energy machine is essentially from the resolved photon where 
any polarization is diluted. In this case and as explained 
at length in the preceding section, laser polarization may be 
sufficient to enhance the signal. The resolved contribution drops 
with lower energies, so that 
at 350~GeV the direct photon contribution, which is critically 
dependent on the polarization, is important. Thus, at the lower energy 
machine one needs as much polarization as possible. 
%
We find that improvements in tagging techniques, better than what 
is achieved with present-day detectors, 
will be advantageous for all the cases that we considered. 
With no such ameliorations and with just 10~fb$^{-1}$ of integrated 
luminosity, it turns out that 
as long as 
a good rejection of the light quarks ({\it i.e.}, other than $c$ and $b$) 
and gluons is provided, all the other parameters for the $b$-tagging 
do not have a critical impact. For instance, at 
both energies, we obtained comparable
results assuming a ``realistic'' tagging efficiency or total
confusion between the $c$ and $b$ quarks. 

With the large value for the significance for the
IMH signal obtained with the
  350~GeV collider, the question naturally arises as to what extent
precision measurements of the $H\gam$ width would be
possible and, more importantly, what could be learnt from them.
Could we indirectly see the effect of new physics
contributing to this one-loop coupling? For instance, the influence of a 
fourth generation, with degenerate doublets (to evade limits from 
$\Delta\rho$), heavy enough not to have been produced directly, has a dramatic 
effect on the two-photon width as shown by 
Gunion and Haber\cite{GunionHaber}. 
Borrowing their illustrative example of quarks weighing $500$~GeV 
and the charged lepton $300$~GeV, 
$\Gamma(H\ra \gam)$ drops to be only 
between 15\% and 30\% of 
its standard model value! The largest drop 
occurs for the lightest Higgs which, 
as we found, are the least easy to extract. 
With this scenario, if by the time this experiment 
is carried out there were no clue about the mass of the 
Higgs
(which may seem unlikely), 
and with so small a width,
there would 
not be enough events to claim discovery of the 
Higgs, let alone to draw any conclusion about 
new physics coupled to the Higgs. 
On the other hand, if the Higgs has already been established and its
mass measured,  then even 
the non-observation of the Higgs in $\gam$ due to a so much smaller 
2-$\gamma$ width
would give precious information on the $H\gam$ coupling and
on the particles that could contribute to it.
As we have discussed, the signal level is directly proportional to  
$\Gamma_{\gam}$ and thus  a variation in the width will be directly
reflected on the significance.  Considering
only statistical errors and assuming the background could
be fitted precisely, the uncertainty on the width
will be 
$\Delta\Gamma/\Gamma \sim ~\sqrt{S+B}/S$.
Therefore, the absolute
number of signal events and the significance
are both  important factors in determining the achievable precision.
We estimate, with an integrated luminosity of 10~fb$^{-1}$,
to be able to  measure the width at the 3$\sigma$ level
with a 
precision of 50\% for $M_H=120$ or $140$~GeV
with the best resolution (5~GeV), polarization and
cuts, while for $M_H=100$~GeV the precision worsen to 84\%.
This might be sufficient though to get some information
on new physics like the 4th generation if the Higgs mass
is known.

We conclude that
the $\gam$ collider obtained from a 500~GeV linear collider 
can discover an  intermediate mass Higgs 
with a luminosity spectrum that allows for other physics studies.
The range 110--140~GeV can be covered with a modest integrated luminosity
of 10~fb$^{-1}$ while the remaining 90--110~GeV calls for higher 
luminosities (about 4 times higher).  
With a lower energy version  of the linear collider (350~GeV),
the whole mass range can be efficiently covered and
precious information on the width of the Higgs into $\gam$
might be gleaned.

\section{Acknowledgements}

We thank our experimental colleagues at LAPP and Universit\'e
de Montr\'eal for numerous 
discussions, in particular, George Azuelos, Dominique Boutigny,
Jean-Pierre Lees and Marie-No\"elle Minard.
We acknowledge helpful discussions  with
Patrick Aurenche et Jean-Philippe Guillet
on QCD issues. We thank Abdel Djouadi for providing us with
the computer code to calculate the Higgs
width.
MB and GB are very grateful to the ENSLAPP theory group for
their warm hospitality.
This work was supported in part by the Natural Science and Engineering
Research Council of Canada.

\end{document}